\documentclass{article}

\usepackage{amsmath,amsfonts,bm}

\newcommand{\EE}{\mathbb{E}}
\newcommand{\norm}[1]{\left\lVert#1\right\rVert}
\newcommand{\cskip}{c_\text{skip}(\sigma)}
\newcommand{\cout}{c_\text{out}(\sigma)}
\newcommand{\cin}{c_\text{in}(\sigma)}
\newcommand{\cnoise}{c_\text{noise}(\sigma)}
\newcommand{\Ft}{F_\theta}
\newcommand{\Dt}{\hat{x}_\theta}

\newcommand{\N}{\mathcal{N}}
\newcommand{\II}{\mathbb{I}}
\newcommand{\R}{\mathbb{R}}
\DeclareRobustCommand{\dataset}[1]{\textsc{\textmd{{#1}}}}

\DeclareMathOperator*{\argmin}{arg\,min}
\DeclareMathOperator{\SVD}{SVD}

\newcommand\numberthis{\addtocounter{equation}{1}\tag{\theequation}}

\newcommand{\score}{\nabla_y \log p_Y}
\newcommand{\denoiser}{\hat{x}}
\newcommand{\noise}{\varepsilon}
\newcommand{\annotate}[2]{%
  {\color{gray}\underbrace{\color{black}#1}_{#2}}%
}
\newcommand{\highlight}[2][1]{%
  \colorlet{highlightcolor}{%
    \ifcase#1\or
      Maroon\or
      RoyalBlue\or
      DarkGreen\or
      Orange\or
      Pink\or
    \else
      Yellow\fi
  }%
  {\color{highlightcolor}{#2}}%
}
\renewcommand{\highlight}[2][1]{#2}

\usepackage{subcaption}
\usepackage{siunitx}
\usepackage{booktabs}
\usepackage{hyperref}
\usepackage{url}
\usepackage{xcolor}
\usepackage{soul}
\usepackage{amsmath}
\usepackage{algorithm}
\usepackage[noend]{algpseudocode}
\usepackage[version=4]{mhchem}
\usepackage{makecell}
\usepackage{wrapfig}
\usepackage{lscape}
\usepackage{rotating}
\usepackage{epstopdf}
\usepackage{verbatim}
\usepackage{enumitem}
\usepackage{microtype}
\usepackage{graphicx}
\usepackage{pifont}
\newcommand{\cmark}{\ding{51}}%
\newcommand{\xmark}{\ding{55}}%

\usepackage{booktabs} 

\newcommand*\Let[2]{\State #1 $\gets$ #2}

\usepackage{hyperref}

\usepackage{arxiv}
\usepackage[numbers]{natbib}

\usepackage{amsmath}
\usepackage{amssymb}
\usepackage{mathtools}
\usepackage{amsthm}

\theoremstyle{plain}

\theoremstyle{definition}

\theoremstyle{remark}

\usepackage[disable,textsize=tiny]{todonotes}

\title{JAMUN: Bridging Smoothed Molecular Dynamics and Score-Based Learning for Conformational Ensembles}
\author{
    Ameya Daigavane\thanks{Equal contribution.} \href{https://orcid.org/0000-0002-5116-3075}{\includegraphics[scale=0.06]{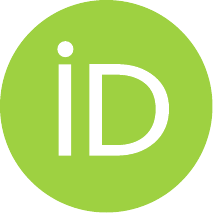}} \\
    Massachusetts Institute of Technology \\
    \texttt{ameyad@mit.edu} \\
    \And
    Bodhi P. Vani\footnotemark[1] \href{https://orcid.org/0000-0002-7747-279X}{\includegraphics[scale=0.06]{orcid.pdf}} \\
    Prescient Design, Genentech \\
    \texttt{vanib@gene.com} \\
    \And
    Darcy Davidson \href{https://orcid.org/0000-0002-4496-2282}{\includegraphics[scale=0.06]{orcid.pdf}} \\
    Prescient Design, Genentech \\
    \texttt{davidsd5@gene.com} \\
    \And
    Saeed Saremi \\
    Prescient Design, Genentech \\
    \texttt{saremis@gene.com} \\
    \And
    Joshua A. Rackers\thanks{Equal contribution.} \href{https://orcid.org/0000-0002-8874-018X}{\includegraphics[scale=0.06]{orcid.pdf}} \\ 
    Prescient Design, Genentech \\
    \And
    Joseph Kleinhenz\footnotemark[2] \href{https://orcid.org/0000-0003-3670-0431}{\includegraphics[scale=0.06]{orcid.pdf}} \\ 
    Prescient Design, Genentech \\
    \texttt{kleinhej@gene.com} \\
}

\hypersetup{
pdftitle={JAMUN: Bridging Smoothed Molecular Dynamics and Score-Based Learning for Conformational Ensembles},
pdfsubject={q-bio.NC, q-bio.QM},
pdfauthor={Ameya Daigavane, Bodhi P. Vani, Darcy Davidson, Saeed Saremi, Joshua A. Rackers, Joseph Kleinhenz},
pdfkeywords={conformational,ensembles,generative,molecular,dynamics,score,smooth},
}

\begin{document}
\maketitle
\begin{abstract}
Conformational ensembles of protein structures are immensely important both for understanding protein function and drug discovery in novel modalities such as cryptic pockets. Current techniques for sampling ensembles such as molecular dynamics (MD) are computationally inefficient, while many recent machine learning methods do not transfer to systems outside their training data. We propose JAMUN which performs MD in a smoothed, noised space of all-atom 3D conformations of molecules by utilizing the framework of walk-jump sampling. JAMUN enables ensemble generation for small peptides at rates of an order of magnitude faster than traditional molecular dynamics.
The physical priors in JAMUN enables transferability to systems outside of its training data, even to peptides that are longer than those originally trained on. Our model, code and weights are available at \url{https://github.com/prescient-design/jamun}.
\end{abstract}

\section{Introduction}
Proteins are inherently dynamic entities constantly in motion, and these movements can be vitally important. They are best characterized as ensembles of structures drawn from the Boltzmann distribution~\citep{Henzler-Wildman2007-cp}, instead of static single structures,  as has traditionally been the case. Protein dynamics is required for the function of most proteins, for instance the global tertiary structure motions for myglobin to bind oxygen and move it around the body~\citep{miller2021moving}, or the beta-sheet transition to a disordered strand for insulin to dissociate and find and bind to its receptor~\citep{insulin}. Similarly, drug discovery on protein kinases depends on characterizing kinase conformational ensembles~\citep{gough2024exploring}. In general the search for druggable `cryptic pockets' requires understanding protein dynamics~\citep{colombo2023computing}, and antibody design is deeply affected by conformational ensembles~\citep{fernandez2023structure}.
However, while machine learning (ML) methods for molecular structure prediction have experienced enormous success recently, ML methods for dynamics have yet to have similar impact.
ML models for generating molecular ensembles are widely considered the `next frontier' \citep{bowman2024alphafold,miller2021moving,zheng2023machine}.
In this work, we present JAMUN (Walk-\textbf{J}ump \textbf{A}ccelerated \textbf{M}olecular ensembles with \textbf{U}niversal \textbf{N}oise), a generative ML model which advances this frontier by demonstrating improvements in both speed and transferability over previous approaches.

While the importance of protein dynamics is well-established, it can be exceedingly difficult to sufficiently sample large biomolecular systems. The most common sampling method is molecular dynamics (MD), which will sample the Boltzmann distribution $p(x) \propto \exp(-U(x))$ defined by the potential $U$ (often called a force field) over all-atom coordinates $\mathcal{X} \subseteq \mathbb{R}^{N \times 3}$, in the limit of infinite sampling time. However, MD is limited by the need for very short timesteps on the order of femtoseconds in the numerical integration scheme; many important dynamical phenomena (such as folding) occur on the much larger timescales of microseconds to milliseconds \citep{folding}. As described by \citet{borhani2012future}, simulating with this resolution is  `\ldots equivalent to tracking the advance and retreat of the glaciers of the last Ice Age -- tens of thousands of years -- by noting their locations each and every second.' Importantly, there is nothing fundamental about this small time-step limitation; it is an artifact of high-frequency motions, such as bond vibrations, that have little effect on protein ensembles \citep{leimkuhler_molecular_2015}. Enhanced sampling methods have been developed in an attempt to accelerate sampling, but they often require domain knowledge about relevant collective variables, and, more importantly, do not address the underlying time-step problem \citep{vitalis2009methods}.  

A large number of generative models have been developed to address the sampling inefficiency problems of MD using machine learning, which we discuss in greater detail in \autoref{sec:baselines}. The key requirement is that of \emph{transferability}: any model must be able to generate conformational ensembles for molecules that are significantly different from those in its training set. To benchmark this transferability, we focus on small peptides whose MD trajectories can be run to convergence in a reasonable amount of time, unlike those for much larger proteins \citep{doi:10.1126/science.1187409}.



\begin{figure}[t]
    \centering
    \includegraphics[width=0.8\linewidth]{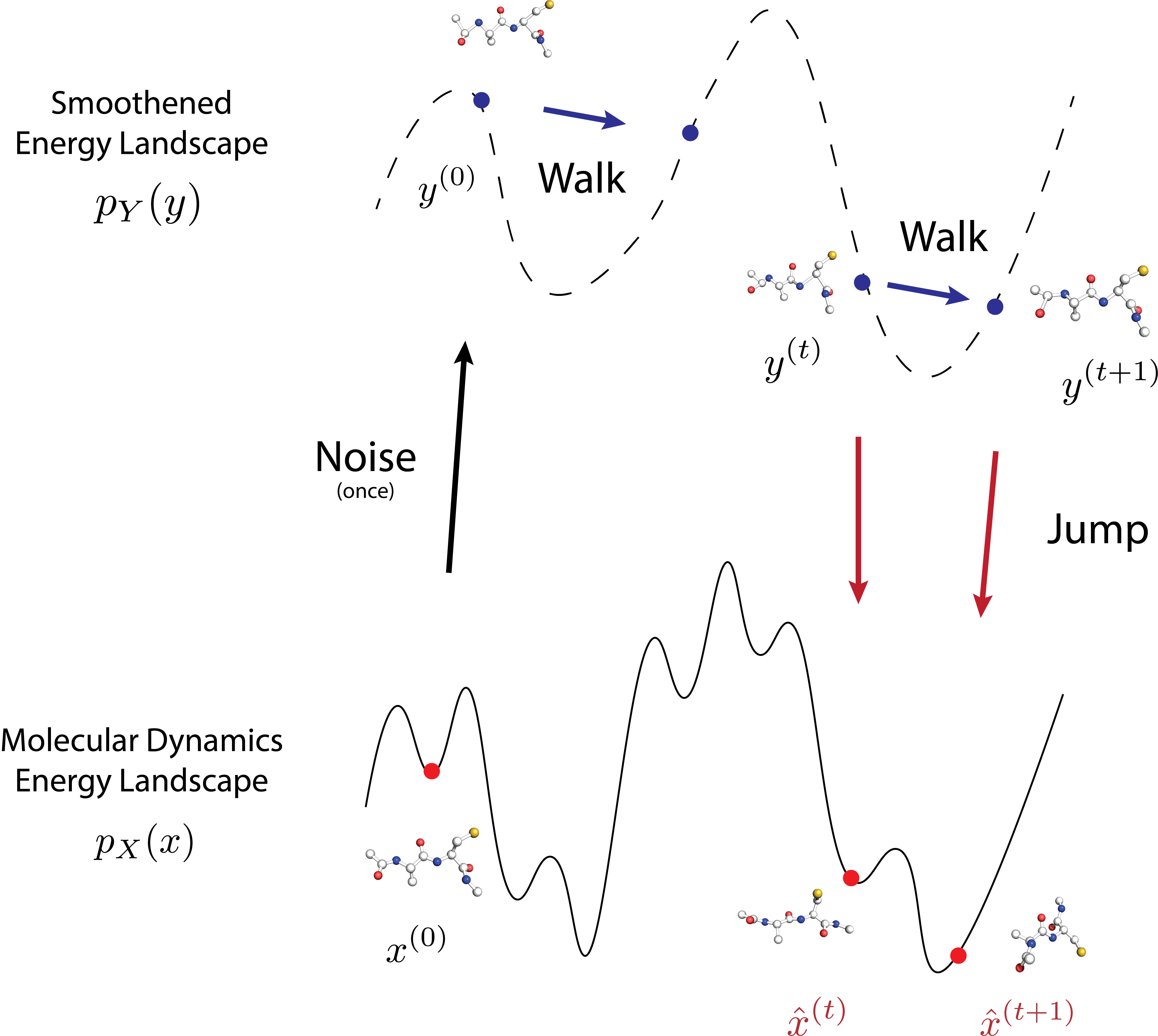}
    \caption{
        Overview of the JAMUN sampling process, where an initial conformation is noised, propagated and denoised to obtain new conformations. 
    }
    \label{fig:walk-jump-overview}
\end{figure}

Here, we propose a new method, JAMUN, that bridges molecular dynamics with score-based learning in a smoothed space. This physical prior enables JAMUN to transfer well -- just like force fields for molecular dynamics  can -- to unseen systems. The key idea is to run Langevin molecular dynamics in a noisy `latent' space $\mathcal{Y} \subseteq \mathbb{R}^{N \times 3}$, instead of the original space $\mathcal{X} \subseteq \mathbb{R}^{N \times 3}$ of all-atom 3D positions. Indeed, $y \in \mathcal{Y}$ is constructed by adding a small amount of independent and identically distributed Gaussian noise $\varepsilon$ to each coordinate in $x \in \mathcal{X}$:
\begin{align}
    y = x + \sigma \varepsilon
\end{align}
To run Langevin MD over $\mathcal{Y}$, the crucial component is the score function $\score(y)$, which needs to be modelled. (This is identical to how a force field needs to be parametrized in classical MD).
Once the MD trajectory over $\mathcal{Y}$ is run, the resulting samples need to be mapped back to clean data via a denoising procedure.

This framework is mathematically described by walk-jump sampling (WJS), as first introduced by \citet{saremi2019neural}. WJS has been used in voxelized molecule generation \citep{pinheiro2024structure,pinheiro2024voxmol} and protein sequence generation \citep{frey2023protein}.
In particular, JAMUN corresponds to a $SE(3)$-equivariant walk-jump sampler of point clouds. As we have described, WJS in this context enjoys several parallels to standard MD.
 
The WJS framework tells us that the score function $\score(y)$ can be used for denoising as well, eliminating the need to learn a separate model. Indeed, a WJS model can be trained similar to a diffusion model \citep{ddpm} with a noising-denoising scheme, but with one key difference: we only need to learn the score function at a single noise level. The choice of this noise level is important; we aim to simply smooth out the distribution enough to resolve sampling difficulties without fully destroying the information present in the data distribution.

In short, the score function $\score(y)$ is learned by adding noise to clean data $x$, and a denoising neural network is trained to recover the clean samples $x$ from $y$. This denoiser defines the score function of the noisy manifold $\mathcal{Y}$ which we sample using Langevin dynamics (walk step) and allows us to periodically project back to the original data distribution (jump step). Crucially, the walk and jump steps are \emph{decoupled} from each other.

Rather than starting over from an uninformative prior for each sample as is commonly done in diffusion \citep{ddpm,ddim} and flow-matching \citep{flow-matching,klein2024equivariant}, JAMUN is able to simply denoise samples from the slightly noised distribution, enabling much greater sampling efficiency. We perform a comparison between full diffusion and walk-jump sampling in \autoref{sec:diffusion-comparison}.

Note that JAMUN is a Boltzmann emulator, unlike a true Boltzmann generator which samples exactly from the Boltzmann distribution defined by $U$.

We train JAMUN on a large dataset of MD simulations of small peptides. We demonstrate that this model can generalize to a holdout set of unseen peptides. In all of these cases, generation with JAMUN yields converged sampling of the conformational ensemble faster than MD with a standard force field, even outperforming several state-of-the-art baselines. 
These results suggest that this transferability is a consequence of retaining the physical priors inherent in MD data. Significantly, we find that JAMUN performs well even for peptides longer than the ones seen in the training set.

Next, we describe the overall working of JAMUN. 

\section{Methods}
\subsection{Representing Peptides as Point Clouds}
\label{sec:point-clouds}

Each point cloud of $N$ atoms can be represented by the tuple $(x, h)$ where $x \in \R^{N \times 3}$ represents the 3D coordinates of each of the $N$ atoms and $h \in \R^{D}$ represents the atom type and covalent bonding information. $h$ can be easily computed from the amino acid sequence for each peptide. We discuss how our model uses $h$ in \autoref{sec:denoiser-parametrization}.
 For clarity of presentation, we omit the conditioning on $h$ in the discussion below.
 
At sampling time, we assume access to an initial sample $x^{(0)} \in \R^{N \times 3}$ sampled from the clean data distribution $p_X$. Similarly to how MD simulations of small peptides are commonly seeded, we use the \texttt{sequence} command in the LEaP program packaged with the Amber force fields to procedurally generate $x^{(0)}$. In theory, $x^{(0)}$ could also be obtained from experimental data, such as crystallized structures from the Protein Data Bank \citep{pdb}. 

\subsection{Walk-Jump Sampling}
\label{sec:walk-jump}

\begin{figure}[ht]
    \centering
    \begin{subfigure}[c]{0.7\linewidth}
        \centering
        \includegraphics[width=\linewidth]{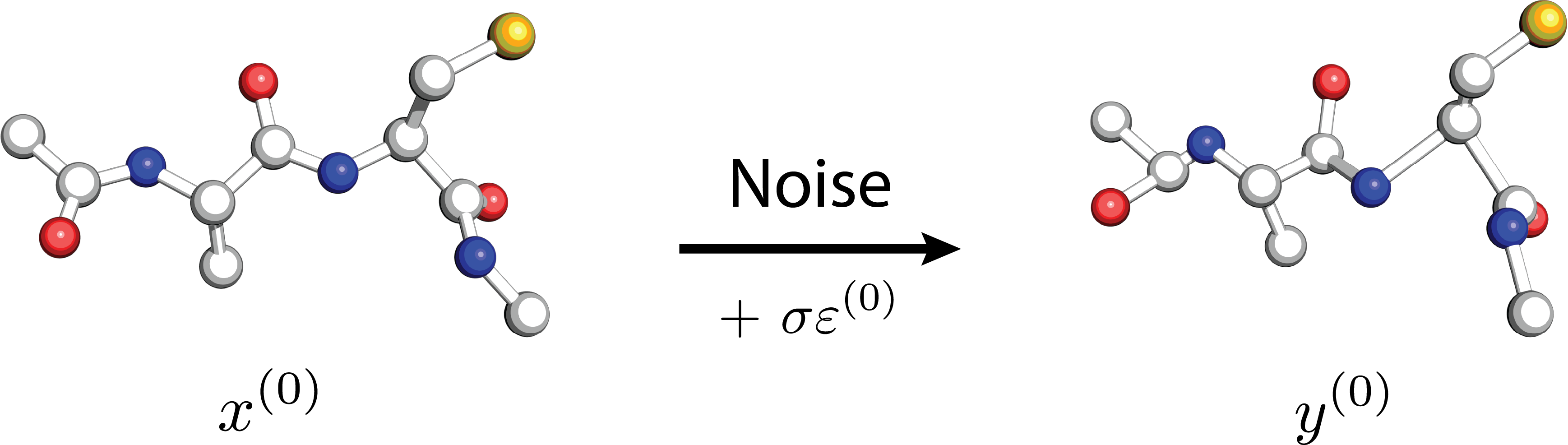}
        \caption{Adding noise to an initial conformation $x^{(0)}$ to obtain $y^{(0)} \sim p_Y$.}
        \label{fig:wj-noise}
    \end{subfigure}\vspace{1em}
    \begin{subfigure}[c]{0.7\linewidth}
        \centering
        \includegraphics[width=\linewidth]{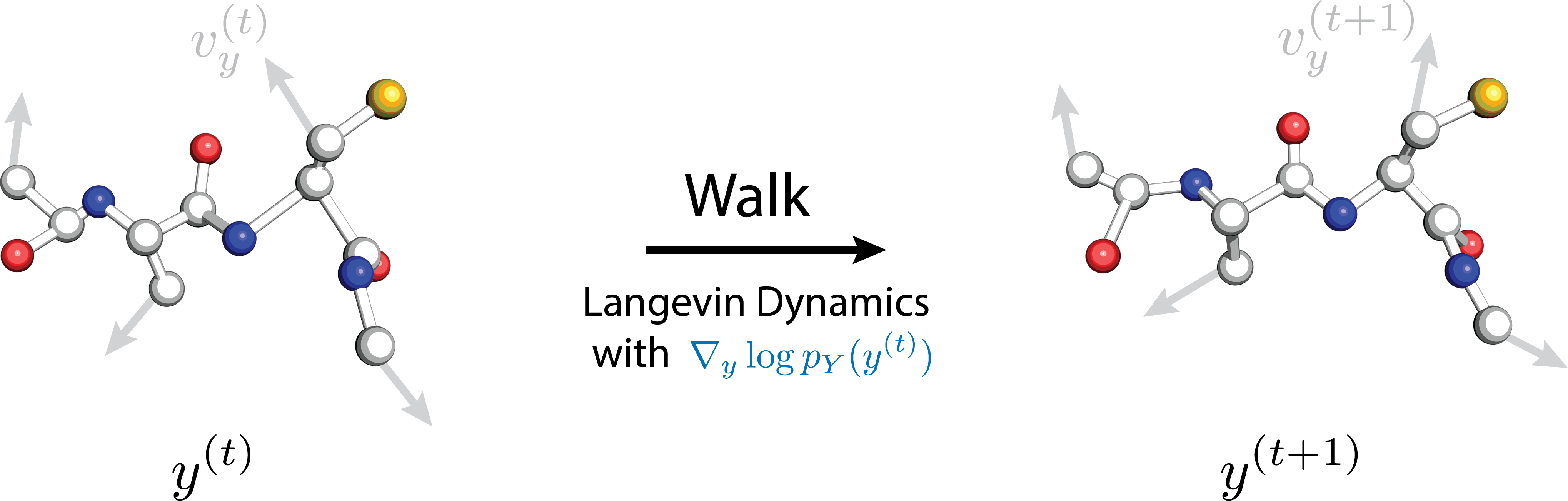}
        \caption{One iteration of BAOAB-discretized Langevin dynamics (\autoref{eqn:langevin} and \autoref{eqn:baoab-update}) starting from $y^{(t)} \sim p_Y$ leads to a new sample $y^{(t + 1)} \sim p_Y$.}
        \label{fig:wj-walk}
    \end{subfigure}\vspace{1em}
    \begin{subfigure}[c]{0.7\linewidth}
        \centering
        \caption{Denoising of $y^{(t)}$ according to \autoref{eqn:score-denoiser} gives us new samples $\highlight[1]{\denoiser^{(t)}}$.}
        \includegraphics[width=\linewidth]{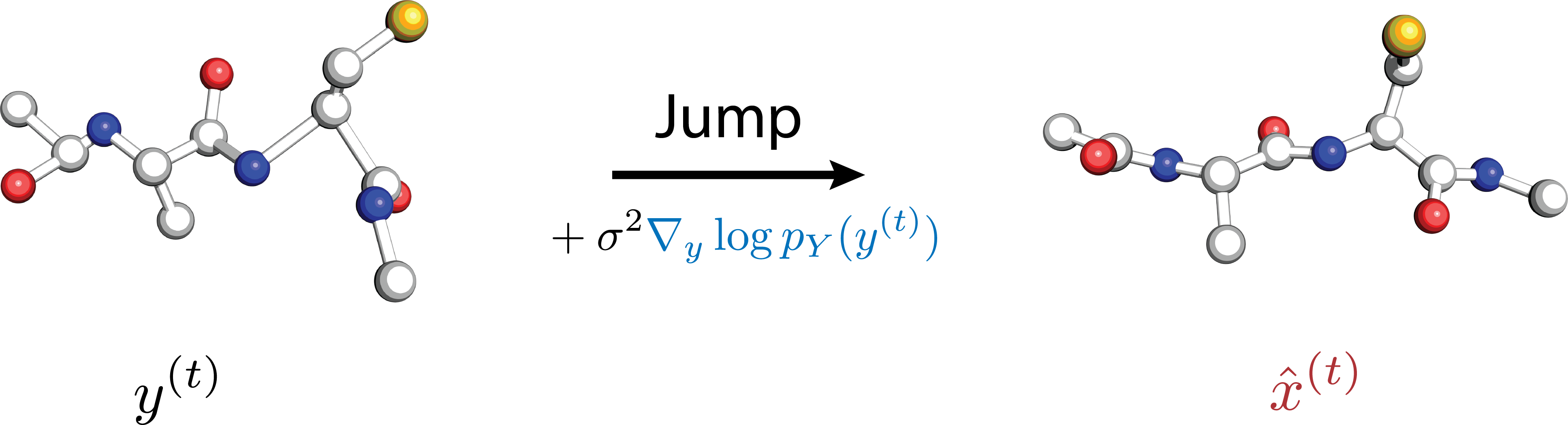}
        \label{fig:wj-jump}
    \end{subfigure}
    \caption{Depictions of the a) initial noising, b) walk and c) jump steps in JAMUN.}

\end{figure}

JAMUN operates by performing walk-jump sampling on molecular systems represented as 3D point clouds.
A conceptual overview of the process is illustrated in \autoref{fig:walk-jump-overview}.

Given the initial sample $x^{(0)} \sim p_X$, walk-jump sampling performs the following steps:
\begin{enumerate}[label=(\alph*)]
    \item \textbf{Noise} the initial structure $x^{(0)}$ to create the initial sample $y^{(0)}$ from the noisy data distribution $p_Y$ (\autoref{fig:wj-noise}):
    \begin{align}
        y^{(0)} &= x^{(0)} + \sigma \noise^{(0)},
        \ \text{where} \  \noise^{(0)} \sim \N(0, \II_{N \times 3}). 
    \end{align}
    
    \item \textbf{Walk} to obtain samples $y^{(1)}, \ldots, y^{(N)}$ from $p_Y$ using Langevin dynamics which conists of numerically solving the following Stochastic Differential Equation (SDE) (\autoref{fig:wj-walk}):
    \begin{align}
        \label{eqn:langevin}
        dy &= v_y dt,
        \\
        dv_y &= \highlight[2]{\score(y)} dt - \gamma v_y dt + M^{-\frac{1}{2}} \sqrt{2} dB_t,
    \end{align}
    where $v_y$ represents the particle velocity, $\highlight[2]{\score(y)}$ is the gradient of the log of the probability density function (called the score function) of $p_Y$, $\gamma$ is friction, $M$ is the mass, and $B_t$ is the standard Wiener process in ${N \times 3}$-dimensions: $B_t \sim \N(0, t \II_{N \times 3})$.
    In practice, we employ the BAOAB solver (\autoref{sec:baoab}) to integrate \autoref{eqn:langevin} numerically.

    \item \textbf{Jump} back to $p_X$ to obtain samples $\denoiser_1, \ldots, \denoiser_N$  (\autoref{fig:wj-jump}) :
    \begin{align}
        \label{eqn:denoiser-definition}
        \denoiser_i = \highlight[1]{\denoiser(y_i)} = \EE[X \ | \ Y = y_i],
    \end{align}
        where $\highlight[1]{\denoiser(\cdot)} \equiv \EE[X \ | \ Y = \cdot ] $ is called the \highlight[1]{denoiser}. It corresponds to the minimizer (\autoref{sec:jump-minimizer-proof}) of the $\ell_2$-loss between clean samples $X$ and samples denoised back from $Y = X + \sigma \noise$.
    \begin{align}
    \label{eqn:jump-minimizer}
        \highlight[1]{\denoiser(\cdot)} = \argmin_{f} \EE_{\substack{X \sim p_X, \noise \sim \N(0, \II_{N \times 3}) \\ Y = X + \sigma \noise}}[\norm{f(Y) - X}^2],         
    \end{align}
where $f: \R^{N \times 3} \to \R^{N \times 3}$.
    As shown by \citet{robbins,miyasawa} (and \autoref{sec:score-denoiser-proof}), the \highlight[1]{denoiser $\denoiser$} is closely linked to the \highlight[2]{score $\score$}:
    \begin{align}
       \label{eqn:score-denoiser}
       \highlight[1]{\denoiser(y)} = y + \sigma^2 \highlight[2]{\score(y)}.
    \end{align}

\end{enumerate}
Importantly, the score function $\highlight[2]{\score}$ shows up in both the \textbf{walk} and \textbf{jump} steps. Next, we show how to approximate the score function to enable walk-jump sampling.

\subsection{Learning to Denoise}
\label{sec:learning-to-denoise}

In order to run walk-jump sampling as outlined above, we have the choice of modelling either the \highlight[2]{score $\score$} or the \highlight[1]{denoiser $\denoiser$} as they are equivalent by \autoref{eqn:score-denoiser}. Following trends in diffusion models \citep{karras2022edm,karras2024edm2}, we model the denoiser as a neural network $\Dt(y, \sigma) \approx \denoiser(y)$ parameterized by model parameters $\theta$. 

Importantly, we only need to learn a model at a \textbf{single, fixed noise level} $\sigma$. This is unlike training diffusion or flow-matching models where a wide range of noise levels are required for sampling. In particular, the choice of noise level $\sigma$ for WJS is important because mode-mixing becomes faster as $\sigma$ is increased, but the task asked of the denoiser becomes harder.

The denoiser $\Dt$ thus takes in noisy point clouds $y$ formed by adding noise (at a fixed noise level $\sigma$) to clean point clouds $x$. The denoiser is tasked to reconstruct back $x$, given $y$. To be precise, training the denoiser $\Dt$ consists of solving the following optimization problem:
\begin{align}
    \label{eqn:denoising-loss}
    \theta^* = \argmin_\theta \EE_{\substack{X \sim p_X, \noise \sim \N(0, \II_{N \times 3}) \\ Y = X + \sigma \noise}}{\norm{{\Dt(Y, \sigma) - X}}^2}
\end{align}
to obtain $\theta^*$, the optimal model parameters. As is standard in the empirical risk minimization (ERM) \citep{erm} setting, we approximate the expectation in \autoref{eqn:denoising-loss} by sampling $X \sim p_X$ and $\noise \sim \N(0, \II_{N \times 3})$. We minimize the loss as a function of model parameters $\theta$ using the first-order optimizer Adam \citep{adam} in PyTorch 2.0 \citep{pytorch2,pytorch-lightning}.

\subsection{Parametrization of the Denoiser Network}
\label{sec:denoiser-parametrization}

We summarize the key features of the denoiser network $\Dt(y, \sigma)$ which will approximate $\denoiser(y)$ in this section. Note that $\sigma$ is fixed in our setting, but we explicitly mention it in this section for clarity.
A diagrammatic overview of our model along with specific hyperparameters are presented in \autoref{sec:denoiser-details}.

We utilize the same parametrization of the denoiser as originally proposed by \citet{karras2022edm,karras2024edm2} in the context of image generation, but with different choices of normalization functions:
\begin{align}
    \label{eqn:denoiser}
    \Dt(y, \sigma) = \cskip y + 
    \cout  \Ft(\cin  y, \cnoise)
\end{align}
where $\Ft$ represents a learned network parameterized by parameters $\theta$.
In particular, $\Ft$ (\autoref{fig:denoiser-message-passing}) is a geometric graph neural network (GNN) model similar to NequIP \citep{nequip,thomas2018tensor}.
Importantly, $\Ft$ is chosen to be $SE(3)$-equivariant, in contrast to existing methods \citep{hoogeboom2022equivariantdiffusionmoleculegeneration,tbg,klein2024timewarp} that utilize the $E(3)$-equivariant EGNN model \citep{satorras2022enequivariantgraphneural}. As rightly pointed out by \citet{dumitrescu2024fieldbasedmoleculegeneration} and \citet{ito}, $E(3)$-equivariant models are equivariant under parity, which means that are forced to transform mirrored structures identically. When we experimented with $E(3)$-equivariant architectures, we found symmetric Ramachandran plots which arise from the unnecessary parity constraint of the denoising network. For this reason, TBG \citep{tbg} and Timewarp \citep{klein2024timewarp} use a `chirality checker' to post-hoc fix the generated structures from their model. For JAMUN, such post-processing is unnecessary because our model can distinguish between chiral structures.

The coefficients
$\cskip, \cout, \cin, \cnoise$ in \autoref{eqn:denoiser} are normalization functions (from $\mathbb{R}^+$ to $\mathbb{R}$) which adjust the effective inputs and outputs to $\Ft$. They are chosen to encourage re-use of the input $y$ at low noise levels, but the opposite at high noise levels. 
Importantly, based on the insight that $\Ft$ uses relative vectors in the message passing steps, we adjust the values of these coefficients instead of simply using the choices made in \citet{karras2022edm,karras2024edm2,boltz1,alphafold3}, as discussed in \autoref{sec:normalization}.

In $\Ft$, edges between atoms are computed using a radial cutoff of $10 \si{\angstrom}$ over the noisy positions in $y$. The edge features are a concatenation of a one-hot feature indicating bonded-ness and the radial distance embedded using Bessel functions. As obtained from $h$, atom-level features are computed using the embedding of the atomic number (eg. $\ce{C}$ and $\ce{N}$), and the atom name following PDB notation (eg. $\ce{CA}, \ce{CB}$ for alpha and beta carbons). Similarly, residue-level features are obtained using the embedding of the residue code (eg. ALA, CYS) and concatenated to each atom in the residue. Importantly, we \emph{do not use the sequence index} of the residues (eg. $0, 1, \ldots$) as we found that it hurts generalization to longer peptide lengths, as has been noted in the context of language modelling \citep{kazemnejad2023impactpositionalencodinglength}.

\section{Datasets}
\label{sec:datasets}

For development, demonstration, and benchmarking against existing models, we use multiple different datasets consisting of peptides from $2$ to $6$ amino acids (AA) long: 
\begin{itemize}
    \item \dataset{Timewarp 2AA-Large} and \dataset{Timewarp 4AA-Large} from \citet{klein2024timewarp},
    \item \dataset{MDGen 4AA-Explicit} from \citet{mdgen2024},
    \item  Our own \dataset{Capped 2AA} and \dataset{Uncapped 5AA} datasets simulated with OpenMM \citep{Openmm_1},
    \item  \dataset{Cremp 4AA} and \dataset{Cremp 6AA} \citep{grambow2024cremp} for conformers of macrocyclic peptides, as computed by the CREST \citep{pracht2020crest} protocol with semi-empirical extended tight-binding (GFN2-xTB) \citep{bannwarth2019gfn2} DFT calculations.
\end{itemize}
The small size of these systems allows us to run converged MD runs to compare to, while the differing simulation conditions across these datasets allow us to test the broad applicability of our approach. Note that such small peptides demonstrate much greater flexibility and conformational diversity than larger proteins, due to a lack of well-defined secondary structure and greater solvent accessibility.
A summary of these datasets is presented in \autoref{tab:datasets}, with a detailed description below.

\begin{figure}[h]
    \centering
    \includegraphics[width=\linewidth]{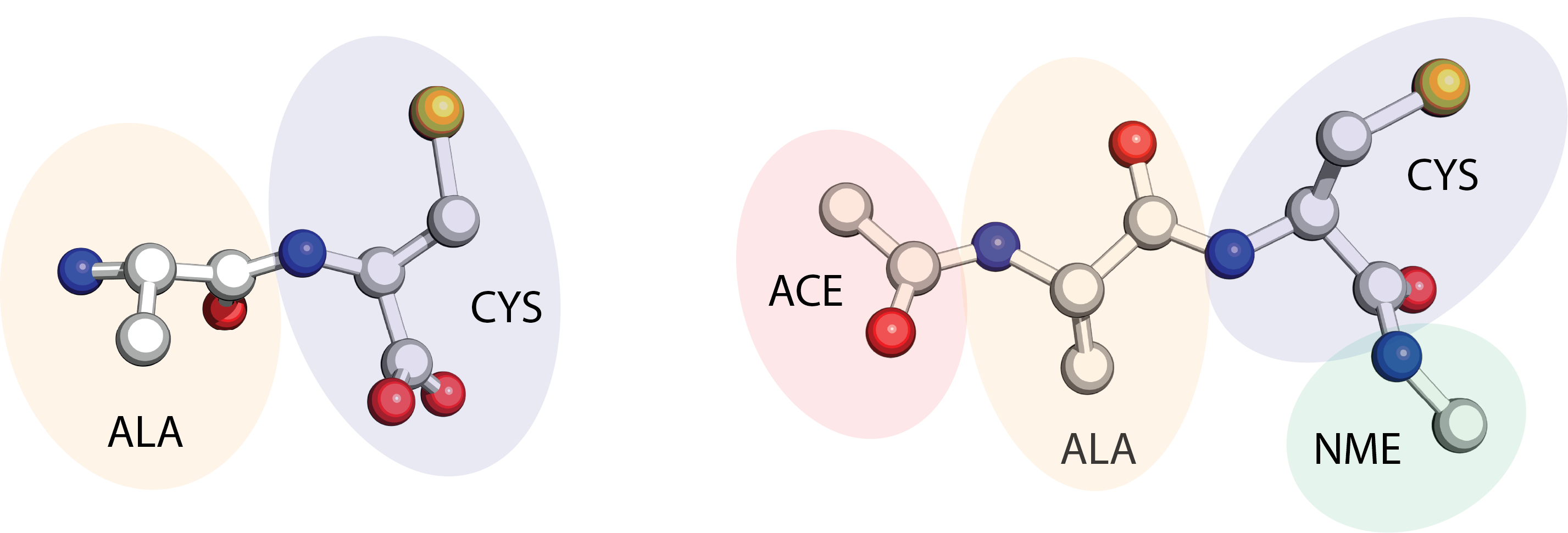}
    \caption{A side-by-side comparison of uncapped (left) compared to capped (right) ALA-CYS. The acetyl (ACE) and N-methyl (NME) capping groups provide steric hindrance and prevent local charge interactions on the N-terminal and C-terminal ends.}
    \label{fig:capped-peptides}
\end{figure}

The \dataset{Timewarp} and \dataset{MDGen} datasets consist of `uncapped' peptides, whose termini are zwitterionic amino and carboxyl groups, as shown in the left panel of \autoref{fig:capped-peptides}. These are not ideal analogues of amino acids in proteins due to local charge interactions as well as lack of steric effects. 

We also create a similar dataset called \dataset{Capped 2AA} of 2AA peptides by adding ACE (acetyl) and NME (N-methyl amide) caps, a common practice in molecular dynamics simulations of very small peptides. As illustrated in the right panel of \autoref{fig:capped-peptides}, these caps introduce additional peptide bonds with the first and last residues. These peptide bonds remove the need for the zwitterion, while the methyl group provides some steric interactions.
These capping groups increase the complexity of the modelling task, but ensure a more realistic distribution of conformations. We choose the same splits as in \dataset{Timewarp 2AA-Large}. Since we simulated this data ourselves, we can also measure the wall-clock speed-ups of JAMUN relative to MD on this dataset. We ensure that our unbiased molecular dynamics runs are converged or representative by comparing against biased molecular dynamics runs using Non-Equilibrium Umbrella Sampling (NEUS) \citep{Dinner2018-ox,Vani2022-ax}, an enhanced sampling algorithm based on trajectory stratification. The simulations are performed at \SI{300}{\K} with the BAOAB integrator \citep{Leimkuhler2013} in OpenMM \citep{Openmm_1}. LINCS \citep{hess1997lincs} is used to constrain the lengths of bonds to hydrogen atoms. Particle Mesh Ewald \citep{PME} is used to calculate electrostatic interactions. The step size was $\SI{2}{fs}$. The systems are solvated with the explicit TIP3P \citep{Jorgensen1983} water model and equilibrated under NVT and NPT ensembles for $\SI{100}{ps}$ each.

For the 2AA datasets, the training set consists of $50\%$ of all possible 2AA peptides. 
For the 4AA datasets, the generalization task is much harder, because the number of 4AA peptides in the training sets is less than $1\%$ and $2\%$ respectively of the total number of possible 4AA peptides.

\begin{table*}[h]
\resizebox{\textwidth}{!}{%
\centering
    \begin{tabular}{ccccccccc}
    \toprule
    Dataset & Peptide Length & Capped? & Force Field & Solvent Model & Temperature & \# Train & \# Validation & \# Test \\  
    \midrule
    \dataset{Timewarp 2AA-Large} &  $2$ & \xmark & \texttt{amber14} & Implicit Water & \SI{310}{\K} & $200$ & $80$ & $100$ \\
    \dataset{Capped 2AA} &  $2$ & \cmark & \texttt{amber99sbildn} & Explicit Water & \SI{300}{\K} & $200$ & $80$ & $100$ \\
    \dataset{Timewarp 4AA-Large} &  $4$ & \xmark & \texttt{amber14} & Implicit Water & \SI{310}{\K} & $1459$ & $379$ & $182$ \\
    \dataset{MDGen 4AA-Explicit} & $4$ & \xmark & \texttt{amber14} & Explicit Water & \SI{350}{\K} & $3109$ & $100$ & $100$ \\
    \dataset{Uncapped 5AA} & $5$ & \xmark & \texttt{amber14} & Implicit Water & \SI{310}{\K} & $-$ & $-$ & $3$ \\
    \dataset{CREMP 4AA} & $4$ & \xmark & \texttt{GFN2-xTB} & Explicit Chloroform & \SI{300}{\K} & $15842$ & $1000$ & $1000$ \\
    \bottomrule
    \end{tabular}
}
\caption{A short description of the simulation conditions across the different datasets.}
\label{tab:datasets}
\end{table*}

We also use the implicit-solvent MD code from Timewarp to generate trajectories for three randomly picked 5AA peptides with codes \texttt{NRLCQ}, \texttt{VWSPF} and \texttt{KTYDI}.  We call this dataset \dataset{Uncapped 5AA}. 

Finally, we investigate conformations of macrocyclic peptides, which are non-linear ring-shape peptides. Macrocyclic peptides are emerging as therapeutic modalities due to their ability to modulate protein-protein interactions \citep{Wang2025} and bind to `undruggable' targets \citep{doi:10.1021/jacs.8b13178}. These molecules present significant challenges in computational modeling because of their conformational diversity and inherent geometric constraints. In fact, sampling their conformations with molecular dynamics is particularly slow, as good classical force-fields are unavailable and it is necessary to use more accurate quantum mechanical energies. Macrocyclic peptides are also extremely unwieldy in their `open', most common conformations, forming hydrogen bond networks with water. However, those macrocycles that are able to occupy smaller `crumpled' conformations are greasy and able to permeate through biological membranes, making them more suitable for biodelivery. We use the macrocycle conformations from CREMP dataset \citep{grambow2024cremp}, specifically their 4AA and 6AA subsets. The CREMP dataset was calculated by the CREST \citep{pracht2020crest} protocol with semi-empirical extended tight-binding (GFN2-xTB) \citep{bannwarth2019gfn2} DFT calculations and the ALPB solvent model in chloroform\cite{Grimme2021}.

\section{Related Work}
\label{sec:baselines}

The goal of building machine learning models that can generate conformational ensembles of molecular systems is not new. While a full overview of this field is beyond the scope of this work -- see \cite{aranganathan2024modeling} for a recent review -- we note a few relevant previous efforts. Boltzmann Generators \citep{noe2019boltzmann} introduced the idea that a neural network could be used to transform the underlying data distribution into an easier-to-sample Gaussian distribution. 
DiffMD \citep{diffmd} learns a diffusion model over conformations of small organic molecules from MD17 \citep{md17}, showing some level of transferability across \ce{C7O2H10} isomers \citep{isomers}.
Timewarp \citep{klein2024timewarp} uses a normalizing flow as a proposal distribution in MCMC sampling of the Boltzmann distribution to approximate the conditional distribution of future conformational states $x^{(t + \Delta t)}$ conditional on the present state $x^{(t)}$. ITO \citep{ito} modelled these transition probabilities using diffusion with a $SE(3)$-equivariant PaiNN architecture. EquiJump \citep{equijump} extended this idea with stochastic interpolants \citep{albergo} and a protein-specific message-passing neural network with reweighting to sample rarer conformations of fast-folding proteins. \citet{hsu2024score} aimed to learn the score of the transition probability distribution to perform `score dynamics', a generalization of molecular dynamics, but still restricted to the original space $\mathcal{X}$.  BoltzNCE \citep{aggarwal2025boltzncelearninglikelihoodsboltzmann} also utilizes stochastic interpolants in a noise contrastive setting to model the conformational space of alanine dipeptide.
Transferable Boltzmann Generators (TBG) \citep{tbg} built upon Timewarp by using flow-matching instead of maximum likelihood estimation and a more efficient continuous normalizing flow architecture, allowing generalization across unseen dipeptides. \citet{kim2024scalablenormalizingflowsenable} also built a scalable Boltzmann generator using a normalizing flow for backbone-only internal coordinates (bond lengths, angles and torsions) demonstrating the ability to sample backbone conformations of the 35-residue HP35 and 56-residue Protein G. Recently, \citet{tan2025scalableequilibriumsamplingsequential} utilized a non-equivariant TarFlow \citep{zhai2025normalizingflowscapablegenerative} architecture along with inference-time Langevin dynamics annealing to improve the scalability of these models. `Two for One' \citep{arts2023two} showed that the score learned by diffusion models can be used for running molecular dynamics simulations.  However, the noise level for the score function is chosen close to $0$, the molecular dynamics is effectively run in the original space $\mathcal{X}$, again limiting the timestep of the simulation.

An important aspect of many of the Boltzmann generator models above is their inability to generalize beyond their training set, with Timewarp and Transferable Boltzmann Generators being notable exceptions. This is important because \emph{only transferable models} can hope to obtain a speed-up over MD in a practical setting; retraining such models is often too expensive.

A related problem is that of protein structure prediction, as tackled by the AlphaFold and related models 
\citep{alphafold1,alphafold2,alphafold3,esmfold,rosettafold}.
Protein structure prediction only requires prediction of a few folded states, and is usually trained on crystallized structures of proteins which can be quite different from their native states. The size of these proteins are significantly larger than the peptides we study here. On the other hand, conformational ensemble generation requires many samples from the Boltzmann distribution to capture the effects of solvent atoms and intramolecular interactions, even if the dynamics is not modelled explicitly. Protein structure prediction models tend to struggle to capture this diversity, especially when sampling the conformations of flexible domains \citep{10.1093/bioinformatics/btac202,fold-switching,af2-beyond-rigid} and rarer conformations \citep{remmebering-af2} not found in the Protein Data Bank (PDB) \citep{pdb}. 
Another issue is that the quality of these models' predictions can be dependent on multiple sequence alignment (MSA) information, a preprocessing step where sequence databases are queried for similar protein sequences which indicate evolutionary conservation patterns as a supplementary input to the model. In fact, some of the first attempts to sample alternative conformations of proteins with AlphaFold2 were performed by subsampling \citep{af2-msa-subsampling}, clustering \citep{af-cluster} or manipulating MSA information \citep{af2-local-energetic-frustration}. Sequence-to-structure models such as ESMFold \citep{esmfold} which do not require MSA information can be faster but tend to produce less physically accurate structures, as noted by \citet{lu2024strstr}.
Unfortunately, MSA information can be quite unreliable for small peptides due to the presence of many hits. In fact, popular MSA software suites such as MMseqs2 \citep{mmseqs2} querying will, by default, simply return an empty MSA for the peptides we study here. 

For a comparison to JAMUN, we use Boltz-1 \citep{boltz1}, an open-source reproduction of AlphaFold3 \citep{alphafold3}. Boltz-1 was trained exclusively on static crystalline structures of folded states, without any dynamics or conformational information, allowing us to evaluate how effectively the conformational landscape can be inferred from structural data alone.

By fixing covalent bond lengths along the backbone and side chains, AlphaFold2 introduced a `frames' parametrization of protein structures consisting of a roto-translation together with $7$ torsion angles for each residue. MDGen \citep{mdgen2024} cleverly builds on this parametrization by creating a $SE(3)$-invariant tokenization of the backbone torsion angles, relative to a known initial conformation (here, $x^{(0)}$). Then, they learn a stochastic interpolant \citep{stoc-interpolant} (a generalization of diffusion and flow-matching) over the trajectories of these tokens. While their overall objective is different from ours -- MD trajectory generation as opposed to Boltzmann distributions -- we can compare to their `forward simulation’ model. Similarly to AlphaFold2, their $SE(3)$-invariant tokenization is limited to single-chain proteins and peptides, but allows for more efficient architectures. On the other hand, JAMUN models all atoms explicitly with a $SE(3)$-equivariant network, which makes it far more flexible and easily extendable to non-linear molecules such as macrocyclic peptides. Importantly, MDGen is a transferable model across the tetrapeptides studied in their original paper.

Several models build upon AlphaFold2 to sample conformational ensembles, which also enables a certain level of transferability due to AlphaFold2's own training strategies. AF2Rank \citep{af2rank} shows that AlphaFold2 can still learn an accurate energy function for protein structures without MSA co-evolution information. \citet{af2-effective-md} finds that AlphaFold2's predicted local distance difference test (pLDDT) and predicted aligned error (PAE) scores correlate with local protein dynamics and global conformational flexibility respectively. They use these scores to parametrize an additional harmonic potential for coarse-grained MD with the MARTINI \citep{martini} force field. AlphaFlow \citep{jing2024alphafold} develops flow-matching over the quotient space of 3D positions modulo rotations to learn a distribution over 3D positions of the $\beta$-carbon atoms, outperforming MSA subsampling with AlphaFold2 over the Protein Data Bank (PDB) \citep{pdb}. 
Distributional Graphormer \citep{zheng2024predicting} parametrizes a diffusion model over $\alpha$-carbon positions, also trained on the PDB. 
Str2Str \citep{lu2024strstr} proposes a noising-denoising process along a range of noise levels for sampling backbone atom coordinates, followed by regression of side-chain coordinates with the rotamer-based FASPR \citep{faspr} packgae. 
BioEmu \citep{bioemu} is a recent backbone-atoms only diffusion model built using the EvoFormer stack from AlphaFold2 \citep{alphafold2}. 
BioEmu is pretrained on $200$ million protein structures from the AlphaFold Protein Structure Database \citep{afdb} and finetuned on over $200 \si{ms}$ of MD data, which are orders of magnitude larger than the datasets we benchmark here. The official repository provides an additional side-chain reconstruction step using H-Packer \citep{hpacker}, allowing comparison to the all-atom models. As seen in \autoref{tab:sampling-times}, side-chain reconstruction can be quite expensive, due to the lack of support for batched inference with H-Packer.

\todo{Add discussion of macrocycle conformation methods, eg. RINGER, RFpeptides}

While much of the work for conformational ensemble generation has focused on linear peptides and proteins, macrocycles -- a class of molecules composed of cyclic arrangements of amino acid residues -- pose unique challenges due to their structural rigidity and conformational intricacies. Physics-based and heuristics-based algorithms such as RDKit's ETKDG \citep{etkdg} and OpenEye's OMEGA \citep{omega_v421} are limited in their ability to predict conformations of macrocyclic peptides. RINGER \citep{grambow2023ringer} and RFpeptides \citep{RFpeptide2025} are two recent machine-learning models for the prediction of macrocycle conformational ensembles. RINGER is a diffusion-based transformer model which encodes their geometry by using redundant internal coordinates. This method accommodates the cyclic nature of macrocycles, as well as stereochemical constraints imposed by sidechains with both L-amino and D-amino acids. RFpeptides uses reinforcement learning to sample the conformational space of macrocyclic systems. By encoding cyclic positional information into its diffusion process, RFpeptides ensures exploration of the structural landscape accessible to macrocycles. However, the reliance on pre-configured scoring functions or domain-specific representations can constrain its broader applicability beyond well-defined cyclic systems. RINGER and RFpeptides are specifically designed and parametrized for macrocyclic peptides: RINGER is tailored for side-chain flexibility and dihedral exploration, while RFpeptides’s cyclic positional encoding restricts its flexibility in adapting to non-macrocyclic topologies. Moreover, these methods incorporate specific heuristics (e.g., redundant internal coordinates) that may not extend efficiently to chemically diverse molecules.

JAMUN addresses many of these limitations by modeling all atoms explicitly with a $SE(3)$-equivariant network, achieving generalization across both linear and non-linear molecules, including macrocyclic peptides. 
\section{Experimental Setup}
\label{sec:setup}

Due to the different simulation conditions across the datasets shown in \autoref{tab:datasets}, we train a different JAMUN model for each dataset. However, the \emph{same noise level} of $\sigma = 0.4 \si{\angstrom}$ is applied for training and sampling on all datasets, highlighting the universality of our approach. In fact, all training hyperparameters \emph{are kept identical} across datasets.
We trained each JAMUN model for $3$ days on $2$ NVIDIA RTX A100 GPUs with $40$ GB memory, although competitive results can be obtained by training for only 1 day. JAMUN is built with the \texttt{e3nn} library \citep{geiger2022e3nn}, and contains approximately $10.5\si{M}$ parameters.

\begin{figure}[h]
    \centering
    \caption{Comparing noise sensitivity for an example test peptide \texttt{GCSL} from \dataset{Timewarp 4AA-Large} for JAMUN, sampled identically, showing the tradeoff between slower mode mixing at $\sigma = 0.2 \si{\angstrom}$, and broken topologies at $\sigma = 0.8 \si{\angstrom}$.}
    \includegraphics[width=0.5\linewidth]{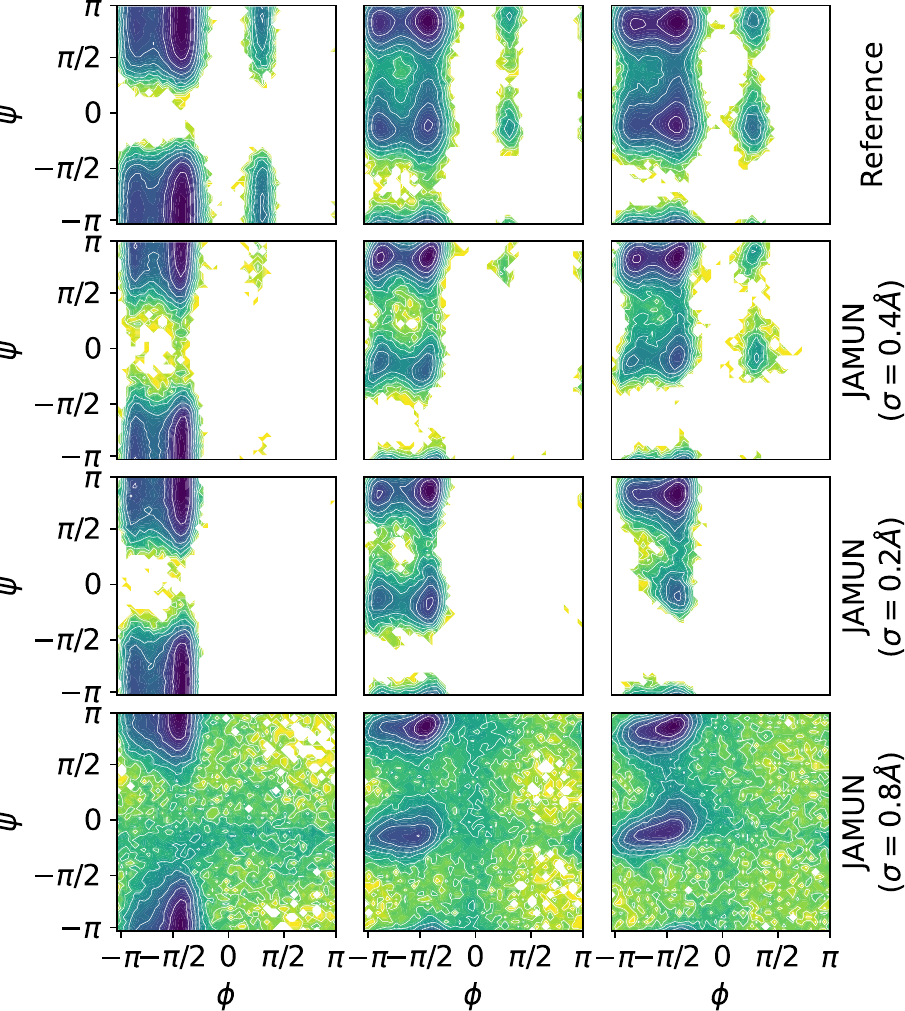}
    \label{fig:noise_comparison}
\end{figure}

This noise scale ($\sigma = 0.4 \si{\angstrom}$) is large enough to result in significant disruption of structure, leading to the smoothed Gaussian convolved `walk' manifold. However, the scale is also small enough to avoid atoms of the same type `swapping' positions, for instance, or pairs of bonded atoms ending up very far from each other with reasonable probability. Indeed, as shown in \autoref{fig:noise_comparison}, we find that higher noise levels (such as $\sigma = 0.8 \si{\angstrom}$) result in samples with broken topologies, while lower noise levels (such as $\sigma = 0.2 \si{\angstrom}$) require many more sampling steps to explore the entire conformational landscape.

\section{Evaluation}
\label{sec:experiments}

Our evaluation aims to answer the following key questions:
\begin{enumerate}
    \item Does JAMUN sample a similar conformational landscape as MD?
    \item Does JAMUN allow for an larger effective timestep than MD, as measured by decorrelation times of sample trajectories?
    \item Does JAMUN enable faster sampling than MD, as measured by wall-clock time?
    \item Can JAMUN generalize to unseen peptides of the same length as seen during training?
    \item Can JAMUN generalize to unseen peptides of the longer length than that seen during training?
\end{enumerate}

\todo{Add decorrelation plots (speedups), MSM state overlaps, transition matrices to show how the sampling paths are different, sampling time tables}

Alongside a comparison to the reference MD simulations, we also compare to several state-of-the-art methods from \autoref{sec:baselines}:
\begin{itemize}
    \item TBG \citep{tbg} on \dataset{Timewarp~2AA-Large}. While the TBG model can technically produce reweighted samples, we run it in the un-reweighted mode to make it a Boltzmann emulator which is almost $10\times$ faster in practice. This allows for a fair comparison to JAMUN. 
    \item MDGen \citep{mdgen2024} on \dataset{MDGen~4AA-Explicit}.
    \item Boltz-1 \citep{boltz1} and BioEmu \citep{bioemu} on \dataset{Uncapped 5AA}.
    \item RINGER \citep{grambow2023ringer} on \dataset{Cremp 4AA} and \dataset{Cremp 6AA}.
\end{itemize} 

To answer these questions, we adopt the analysis methods and metrics from MDGen \citep{mdgen2024}. In particular, we project sampled and reference distributions of all-atom positions onto a variety of variables: pairwise distances, dihedral angles of backbone (known as Ramachandran plots) and sidechain torsion angles, TICA \citep{tica} (time-lagged independent coordinate analysis) projections, and metastable state probabilities as computed by Markov State Models (MSMs) fit with PyEMMA \citep{pyemma2}. TICA is a popular dimensionality reduction method for larger molecules which aims to extract slow collective degrees of freedom from a trajectory \citep{tica-identification,tica-non-native}. 
As is standard practice, all TICA projections and MSMs are estimated using the reference MD data.

\begin{table*}[h]
    \caption{Comparison of (approximate and batched) sampling times per test peptide for baseline models (top), baseline MD simulations (middle) and JAMUN. All models and baselines were run on a single NVIDIA RTX A100 GPU, except for \dataset{MDGen 4AA-Explicit} which was simulated on a single NVIDIA T4 GPU, as mentioned in \citet{mdgen2024}. Sampling times for the original \dataset{Timewarp} datasets are unknown.}
    \centering
    \resizebox{0.8\linewidth}{!}{%
        \begin{tabular}{cccc}
            \toprule
            Model & \makecell{Time per Sample} & Number of Samples & \makecell{Total Time}\\  
            \midrule
            JAMUN (2AA) & \SI{2}{\milli\second} & $100,000$ & \SI{3}{\minute} \\
            JAMUN (4AA) & \SI{3}{\milli\second} & $100,000$ & \SI{5}{\minute} \\
            JAMUN (5AA) & \SI{8}{\milli\second} & $100,000$ & \SI{12.5}{\minute} \\
            JAMUN (4AA Macrocycle) & \SI{2.4}{\milli\second} & $100,000$ & \SI{4}{\minute} \\
            \midrule
            TBG  & \SI{720}{\milli\second} & $5,000$ & \SI{60}{\minute} \\
            MDGen & \SI{6}{\milli\second} & $10,000$ & \SI{1}{\minute} \\
            Boltz-1  & \SI{360}{\milli\second} & $10,000$ & \SI{60}{\minute} \\
            BioEmu & \SI{15}{\milli\second} & $10,000$ & \SI{2.5}{\minute} \\
            BioEmu with H-Packer & \SI{4320}{\milli\second} & $10,000$ & \SI{720}{\minute} \\
            RINGER  & \SI{1.3}{\milli\second} & $5000$ & \SI{6}{\second} \\
            RINGER with side-chain reconstruction & \SI{111}{\milli\second} & $5000$ & \SI{9.5}{\minute} \\
            \midrule
            \dataset{Capped 2AA} & \SI{40}{\milli\second} & $60,000$ & \SI{40}{\minute} \\
            \dataset{MDGen 4AA-explicit} & \SI{11}{\milli\second} & $1,000,000$ & \SI{180}{\minute} \\
            \dataset{Uncapped 5AA} & \SI{108}{\milli\second} & $100,000$ & \SI{180}{\minute} \\
            \dataset{Cremp 4AA} & \SI{156}{\milli\second} & $100,000$ & \SI{258}{\hour} \\
            \bottomrule
        \end{tabular}
    }
    \label{tab:sampling-times}
\end{table*}

\autoref{tab:sampling-times} contains a summary of the sampling efficiencies for different models and baseline methods, averaged over the \emph{unseen} peptides in the corresponding test set.

\subsection{JAMUN samples a similar conformational landscape as MD}
\label{sec:metastable_states}

\begin{figure}[h!]
    \centering
    \includegraphics[width=0.24\linewidth]{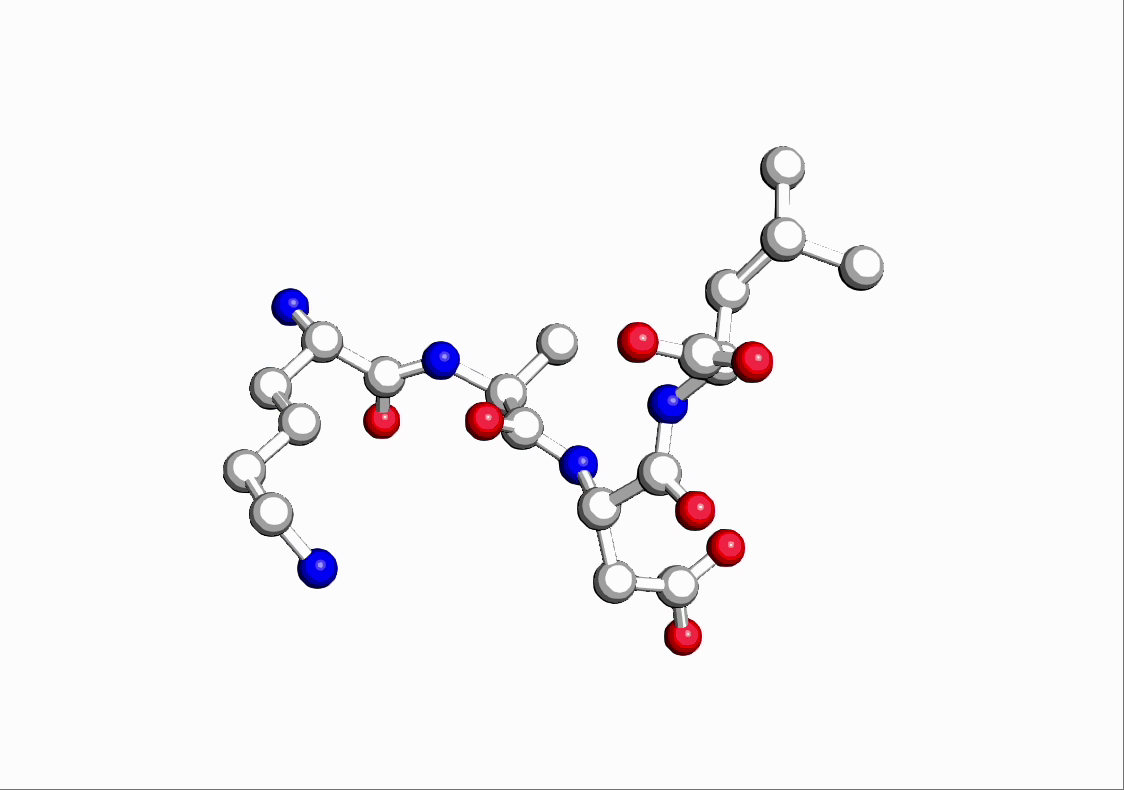}
    \includegraphics[width=0.24\linewidth]{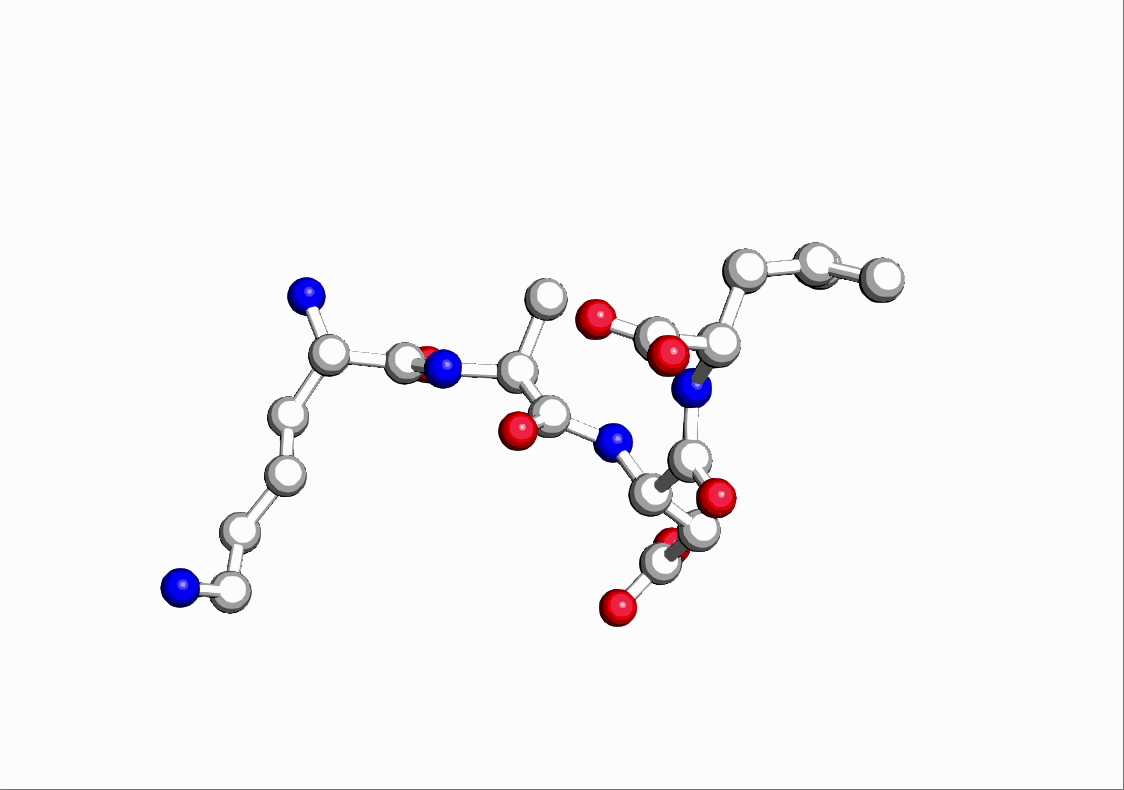}
    \includegraphics[width=0.24\linewidth]{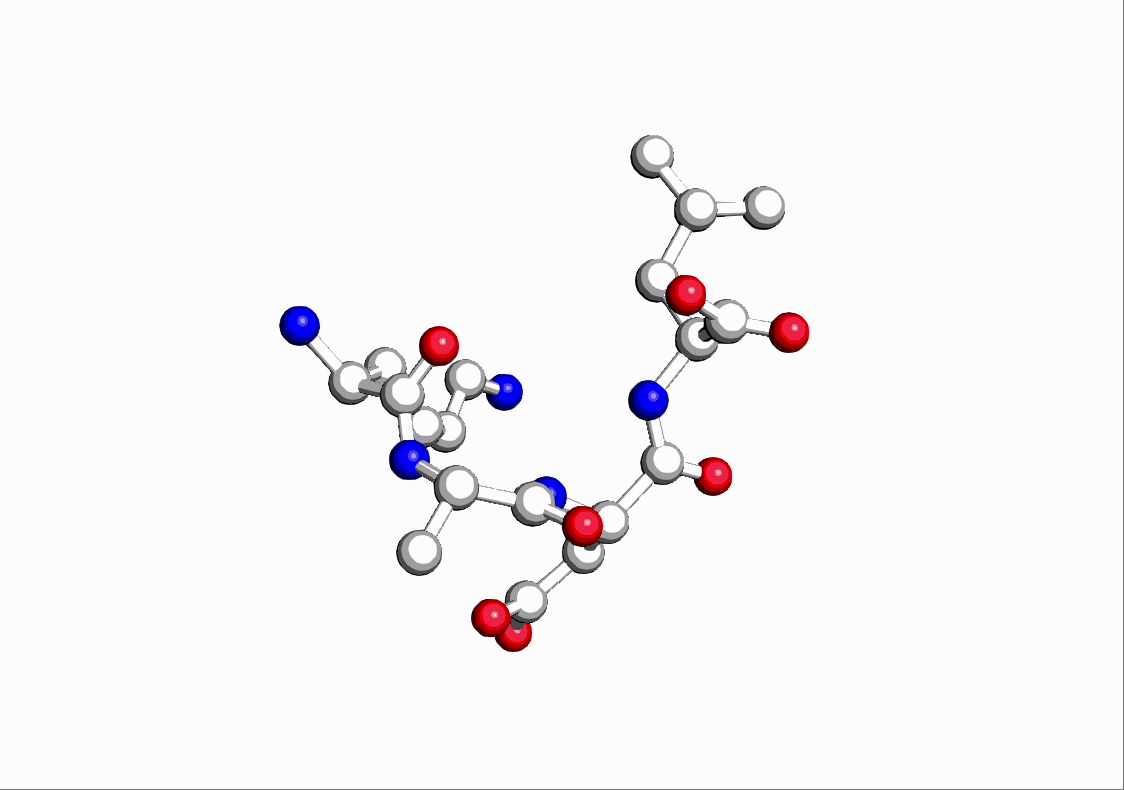}
    \includegraphics[width=0.24\linewidth]{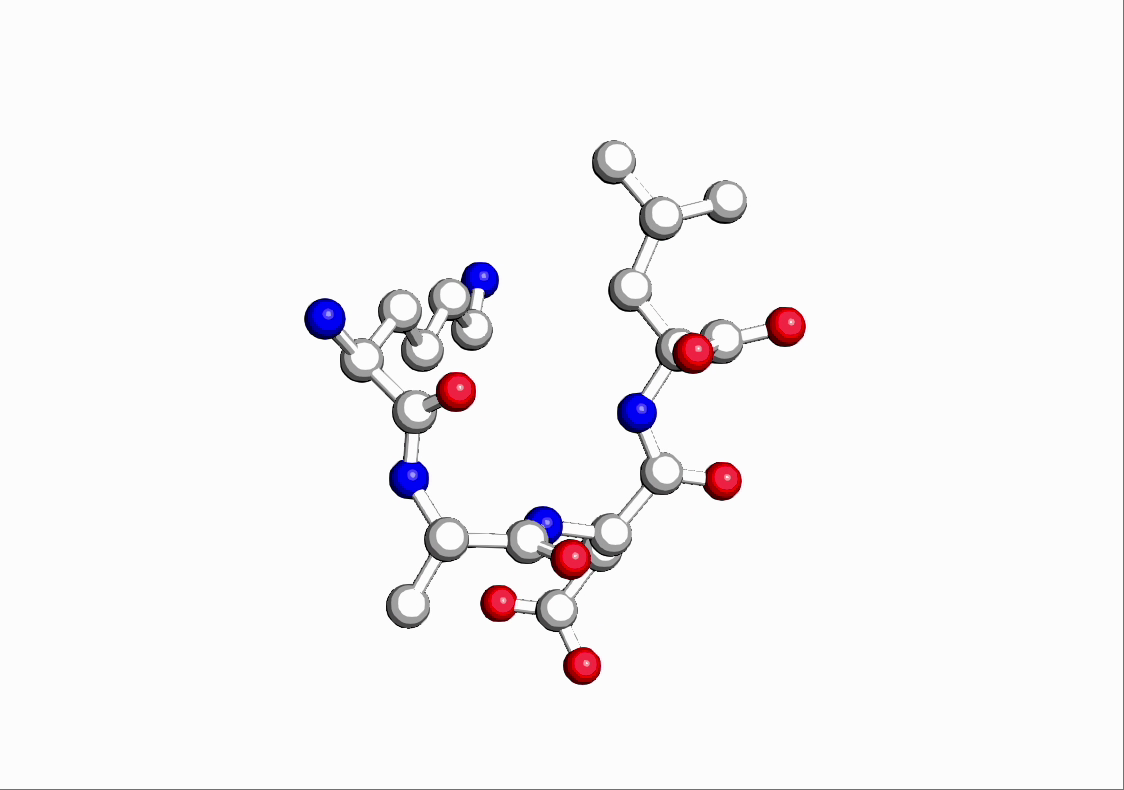}

    \caption{JAMUN samples on unseen \texttt{KADL} when trained on \dataset{Timewarp 4AA-Large}. The full animation is available at \url{https://github.com/prescient-design/jamun}.}
    \label{fig:jamun-samples}
\end{figure}

\begin{figure}[h]
    \centering
    \caption{Across the  \dataset{Timewarp 2AA-Large} (left), \dataset{Timewarp 4AA-Large} (middle)  and \dataset{MDGen 4AA-Explicit} (right) datasets, MSM state probabilities for JAMUN samples (on the $y$-axis) and those for the reference MD trajectories (on the $x$-axis) across all test peptides are strongly correlated. The perfect sampler will obtain an $R^2$ of $1$.}
\includegraphics[width=0.3\linewidth]{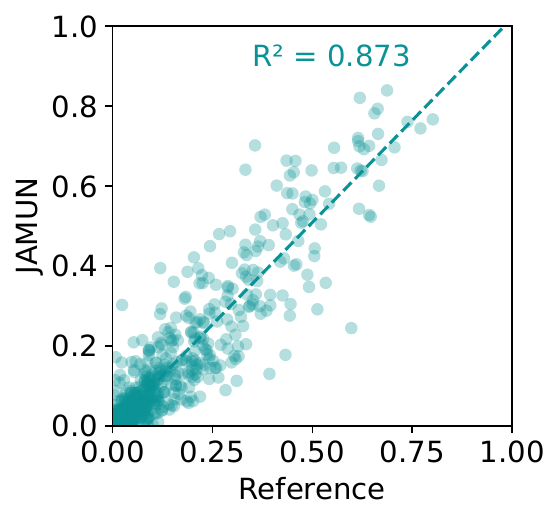}
\includegraphics[width=0.3\linewidth]{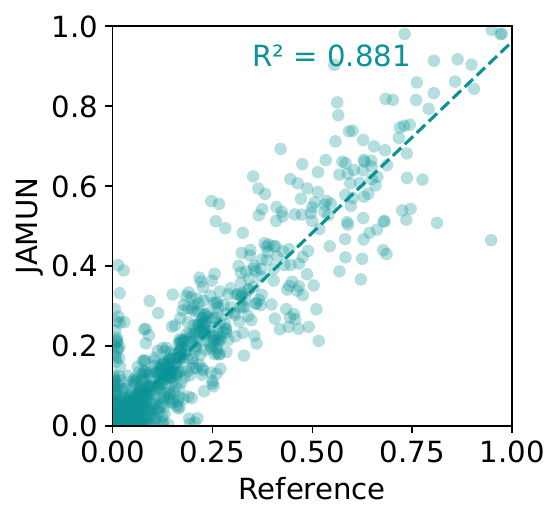}
\includegraphics[width=0.3\linewidth]{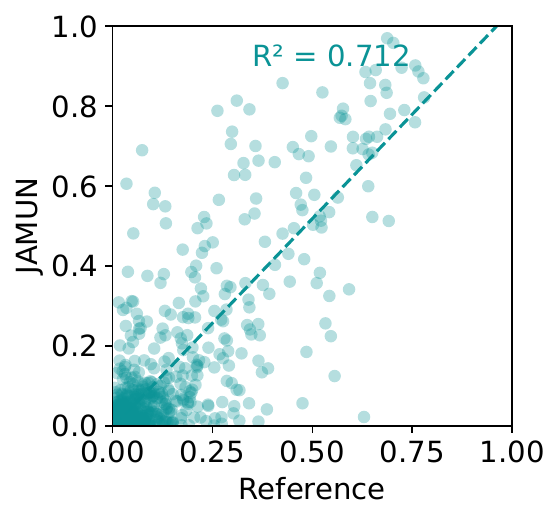}
\label{fig:metastable_probs}
\end{figure}

As visually suggested in \autoref{fig:Timewarp_2AA_ramachandrans} and \autoref{fig:Timewarp_2AA_tica_projections}, JAMUN samples similar states to those in the reference MD data. Indeed, as \autoref{fig:metastable_probs} collected over all \emph{unseen} test peptides shows, the metastable state probabilities over JAMUN sampled trajectories match very well with those over the reference MD data on the \dataset{Timewarp 2AA-Large} and \dataset{Timewarp 4AA-Large} datasets.

We perform an analysis of the physical validity and the energy of JAMUN samples on \dataset{MDGen 4AA-Explicit} in \autoref{sec:physical-validity}. In summary, we find that JAMUN samples pass all Posebusters \citep{posebusters} checks at a rate of $\approx 94.7\%$ and have energies close to the ground truth MD for randomly selected $20$ unseen peptides.

\subsection{JAMUN decorrelates faster than MD}
\label{sec:decorrelation}

\begin{figure}[h]
    \centering
    \includegraphics[width=0.4\linewidth]{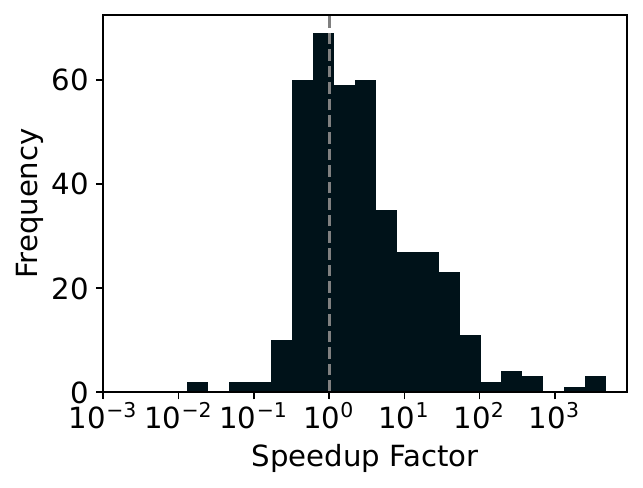}
    \includegraphics[width=0.4\linewidth]{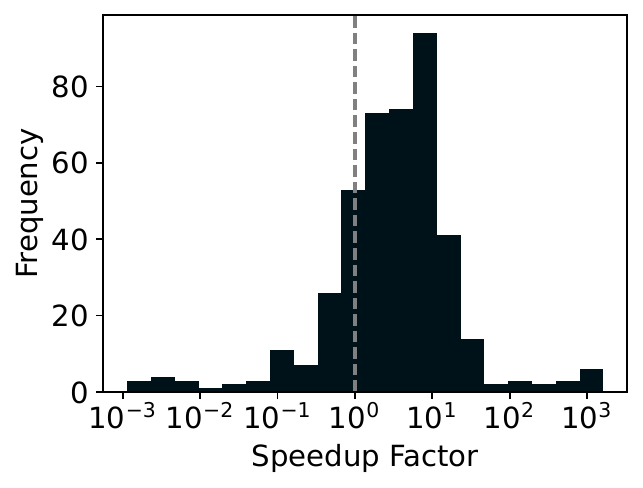}
    \caption{Speedups defined as the ratio between decorrelation times between the reference MD and JAMUN for backbone (left) and sidechain (right) torsions for all test peptides in \dataset{Capped 2AA}.}
\label{fig:Capped_2AA_torsion_decorrelation_times}
\end{figure}

\begin{figure}[h]
    \centering
    \includegraphics[width=0.4\linewidth]{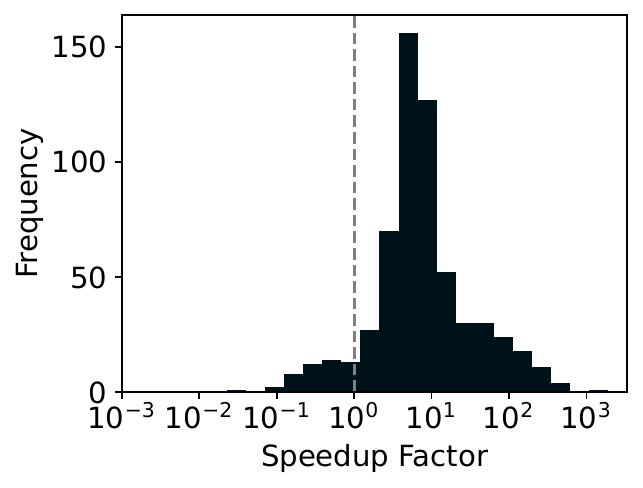}
    \includegraphics[width=0.4\linewidth]{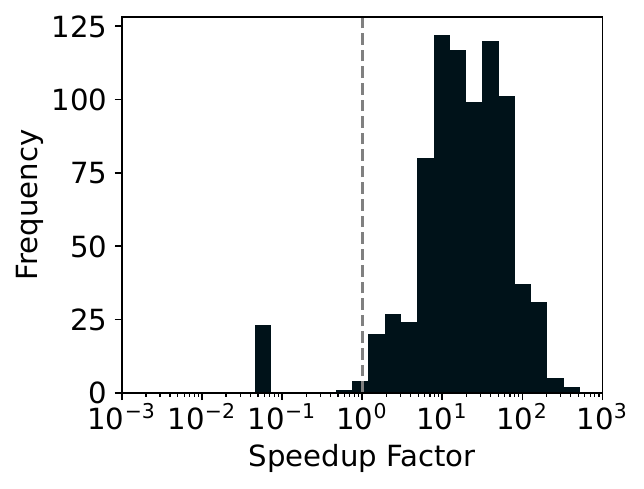}
    \caption{Speedup (ratio of decorrelation time) for backbone (left) and sidechain (right) torsions, histogrammed for all test peptides from the \dataset{MDGen 4AA-Explicit} dataset.}
\label{fig:MDGen_4AA_speedups_combined}
\end{figure}

Here, we compute the ratio of the decorrelation times for the backbone and sidechain torsions in JAMUN and the reference MD data. \autoref{fig:Capped_2AA_torsion_decorrelation_times} and \autoref{fig:MDGen_4AA_speedups_combined}
highlight how sampling in the smoothed space $\mathcal{Y}$ compared to the original space $\mathcal{X}$ enables much faster decorrelation due to the ability to take larger `steps'.

In \autoref{tab:Capped_2AA_jsd_table}, we compare JAMUN with the reference MD, shortened by a factor of $10$, using the Jensen-Shannon distance to the full reference MD data. JAMUN outperforms this shortened MD trajectory across all metrics; even though the shortened MD trajectory takes approximately $2\times$ longer to sample than JAMUN, from \autoref{tab:sampling-times}.

We did not use the \dataset{Timewarp} datasets for this analysis, because they seem to be concatenations of subsampled MD trajectories and not a single MD trajectory. Thus, their true decorrelation times cannot be estimated.

\subsection{JAMUN enables faster (approximate) sampling than MD}
From the results in \autoref{sec:metastable_states}, \autoref{sec:decorrelation} and
\autoref{tab:sampling-times}, we would be inclined to say that JAMUN leads to a significant speedup in wall-clock time over MD. 

However, we would like to highlight here that JAMUN does not provide reweighted samples, nor does it come with any guarantees (finite or asymptotic) about the convergence of the sampled distribution. This can be dealt with by learning an energy model $U_\mathcal{Y}$ to parametrize the unnormalized   distribution $p_\mathcal{Y}$ directly, which allows one to compute the reweighting factors for importance sampling. We leave this extension for future work. 

\begin{figure}[h]
    \centering
    \includegraphics[width=0.95\linewidth]{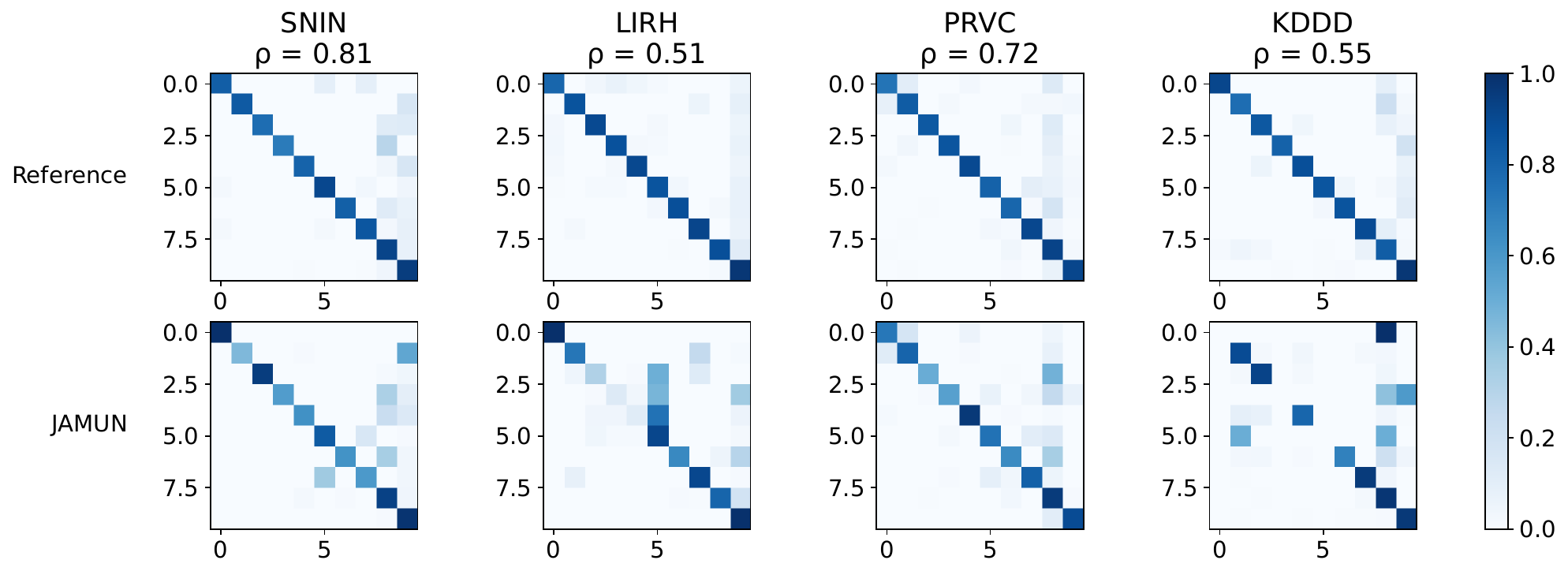}
    \caption{Transition matrices between metastable states for $4$ randomly picked test peptides in the \dataset{MDGen 4AA-Explicit} dataset. $\rho \in [0, 1]$ indicates the Spearman correlation between the reference MD and JAMUN transition matrices.}
\label{fig:MDGen_4AA_transition_matrices}
\end{figure}

Note that the JAMUN dynamics (and hence the time step) in $\mathcal{Y}$ cannot be mapped to the dynamics of the original MD in $\mathcal{X}$, because a single configuration $y$ can arise from the noising of many $x$. In practice, as seen in \autoref{fig:MDGen_4AA_transition_matrices}, the empirical transition matrices between metastable states look quite different between JAMUN and the reference MD.

In conclusion, we can say that JAMUN allows the approximate sampling of the conformational landscape with significant speed-up over MD. Clearly, the speedup is larger when explicit solvent is used (such as in \dataset{Capped 2AA}), but is also evident in implicit solvent (such as in \dataset{Uncapped 5AA}).

\begin{table*}[h]
\caption{Comparison of Jenson-Shannon distances between JAMUN and the reference MD (shortened by a factor of $10$), averaged over the test peptides in \dataset{Capped 2AA}. Note that this shortened reference MD takes $2 \times$ longer to sample as JAMUN.}
\resizebox{\columnwidth}{!}{%
\centering
    \begin{tabular}{lcccccc}
    \toprule
    Trajectory & \makecell{Backbone\\Torsions} & \makecell{Sidechain\\Torsions} & \makecell{All\\Torsions} & TICA-0 & \makecell{TICA-0,1} & \makecell{Metastable\\Probs} \\  
    \midrule
    JAMUN & $0.291 \pm 0.119$ & $0.320 \pm 0.108$ & $0.304 \pm 0.112$ & $0.351 \pm 0.130$ & $0.438 \pm 0.117$ & $0.264 \pm 0.108$ \\
    Reference ($10 \times$ shorter) & $0.447 \pm 0.057$ & $0.406 \pm 0.071$ & $0.424 \pm 0.056$ & $0.557 \pm 0.043$ & $0.564 \pm 0.041$ & $0.543 \pm 0.073$ \\
    \bottomrule
    \end{tabular}
}
\label{tab:Capped_2AA_jsd_table}
\end{table*}

\begin{table*}[h]
\centering
\caption{Comparison of Jenson-Shannon distances between JAMUN and TBG for \dataset{Timewarp 2AA-Large}. Note that TBG ($20 \times$ shorter) has a similar sampling time as JAMUN.}
\label{tab:Timewarp_2AA_jsd_table}
\resizebox{\columnwidth}{!}{%
    \begin{tabular}{lcccccc}
        \toprule
        Trajectory & \makecell{Backbone \\ Torsions} & \makecell{Sidechain \\ Torsions} & \makecell{All \\ Torsions} & TICA-0 & \makecell{TICA-0,1} & \makecell{Metastable \\ Probs} \\
        \midrule
        JAMUN & $0.130 \pm 0.020$ & $0.185 \pm 0.044$ & $0.165 \pm 0.030$ & $0.177 \pm 0.053$ & $0.260 \pm 0.052$ & $0.155 \pm 0.063$ \\
        TBG & $0.083 \pm 0.028$ & $0.115 \pm 0.045$ & $0.105 \pm 0.038$ & $0.122 \pm 0.051$ & $0.225 \pm 0.070$ & $0.101 \pm 0.046$ \\
        TBG ($20 \times$ shorter) & $0.203 \pm 0.070$ & $0.235 \pm 0.073$ & $0.225 \pm 0.071$ & $0.240 \pm 0.072$ & $0.484 \pm 0.073$ & $0.124 \pm 0.054$ \\
        \bottomrule
    \end{tabular}%
}
\end{table*}

\begin{table*}[h]
\centering
\caption{Comparison of Jenson-Shannon distances between JAMUN, MDGen, and MD trajectories (shortened by a factor of $10$ and $100$) for the \dataset{MDGen 4AA-Explicit}.}
\label{tab:MDGen_4AA_jsd_table}
\resizebox{\columnwidth}{!}{%
    \begin{tabular}{lcccccc}
        \toprule
        Trajectory & \makecell{Backbone \\ Torsions} & \makecell{Sidechain \\ Torsions} & \makecell{All \\ Torsions} & TICA-0 & \makecell{TICA-0,1} & \makecell{Metastable \\ Probs} \\
        \midrule
        JAMUN & $0.159 \pm 0.060$ & $0.210 \pm 0.057$ & $0.187 \pm 0.054$ & $0.257 \pm 0.111$ & $0.353 \pm 0.120$ & $0.262 \pm 0.118$ \\
        Ref. MD ($10 \times$ shorter) & $0.100 \pm 0.035$ & $0.092 \pm 0.027$ & $0.095 \pm 0.025$ & $0.234 \pm 0.068$ & $0.332 \pm 0.067$ & $0.286 \pm 0.066$ \\
        Ref. MD ($100 \times$ shorter) & $0.227 \pm 0.062$ & $0.254 \pm 0.060$ & $0.240 \pm 0.051$ & $0.444 \pm 0.131$ & $0.569 \pm 0.108$ & $0.482 \pm 0.138$ \\
        MDGen & $0.129 \pm 0.039$ & $0.089 \pm 0.032$ & $0.107 \pm 0.028$ & $0.228 \pm 0.092$ & $0.320 \pm 0.087$ & $0.233 \pm 0.093$ \\
        \bottomrule
    \end{tabular}%
}
\end{table*}

\begin{table*}[h]
\centering
\caption{Comparison of Jenson-Shannon distances between JAMUN, Boltz-1, BioEmu, and reference MD (shortened by a factor of $10$ and $100$), for three test peptides in \dataset{Uncapped 5AA}.}
\label{tab:Uncapped_5AA_jsd_table}
\resizebox{\columnwidth}{!}{%
    \begin{tabular}{lcccccc}
        \toprule
        Trajectory & \makecell{Backbone \\ Torsions} & \makecell{Sidechain \\ Torsions} & \makecell{All \\ Torsions} & TICA-0 & \makecell{TICA-0,1} & \makecell{Metastable \\ Probs} \\
        \midrule
        JAMUN & $0.196 \pm 0.027$ & $0.196 \pm 0.013$ & $0.197 \pm 0.010$ & $0.336 \pm 0.049$ & $0.440 \pm 0.048$ & $0.250 \pm 0.075$ \\
        Ref. MD ($10 \times$ shorter) & $0.118 \pm 0.013$ & $0.150 \pm 0.032$ & $0.135 \pm 0.015$ & $0.430 \pm 0.077$ & $0.504 \pm 0.079$ & $0.460 \pm 0.051$ \\
        Ref. MD ($100 \times$ shorter) & $0.272 \pm 0.062$ & $0.307 \pm 0.023$ & $0.290 \pm 0.039$ & $0.555 \pm 0.070$ & $0.678 \pm 0.034$ & $0.601 \pm 0.112$ \\
        Boltz-1 & $0.425 \pm 0.033$ & $0.402 \pm 0.036$ & $0.411 \pm 0.029$ & $0.457 \pm 0.050$ & $0.584 \pm 0.026$ & $0.483 \pm 0.047$ \\
        BioEmu & $0.329 \pm 0.013$ & $0.489 \pm 0.024$ & $0.420 \pm 0.018$ & $0.415 \pm 0.092$ & $0.597 \pm 0.026$ & $0.321 \pm 0.018$ \\
        \bottomrule
    \end{tabular}%
}
\end{table*}

\subsection{Comparison to TBG and MDGen}

\autoref{tab:Timewarp_2AA_jsd_table} shows that on \dataset{Timewarp 2AA-Large}, JAMUN outperforms TBG when run for equal amounts of time (based on \autoref{tab:sampling-times}), and is only slightly worse when TBG is run for $20\times$ longer.

\begin{figure}[h!]
    \centering
    \includegraphics[width=0.7\linewidth]{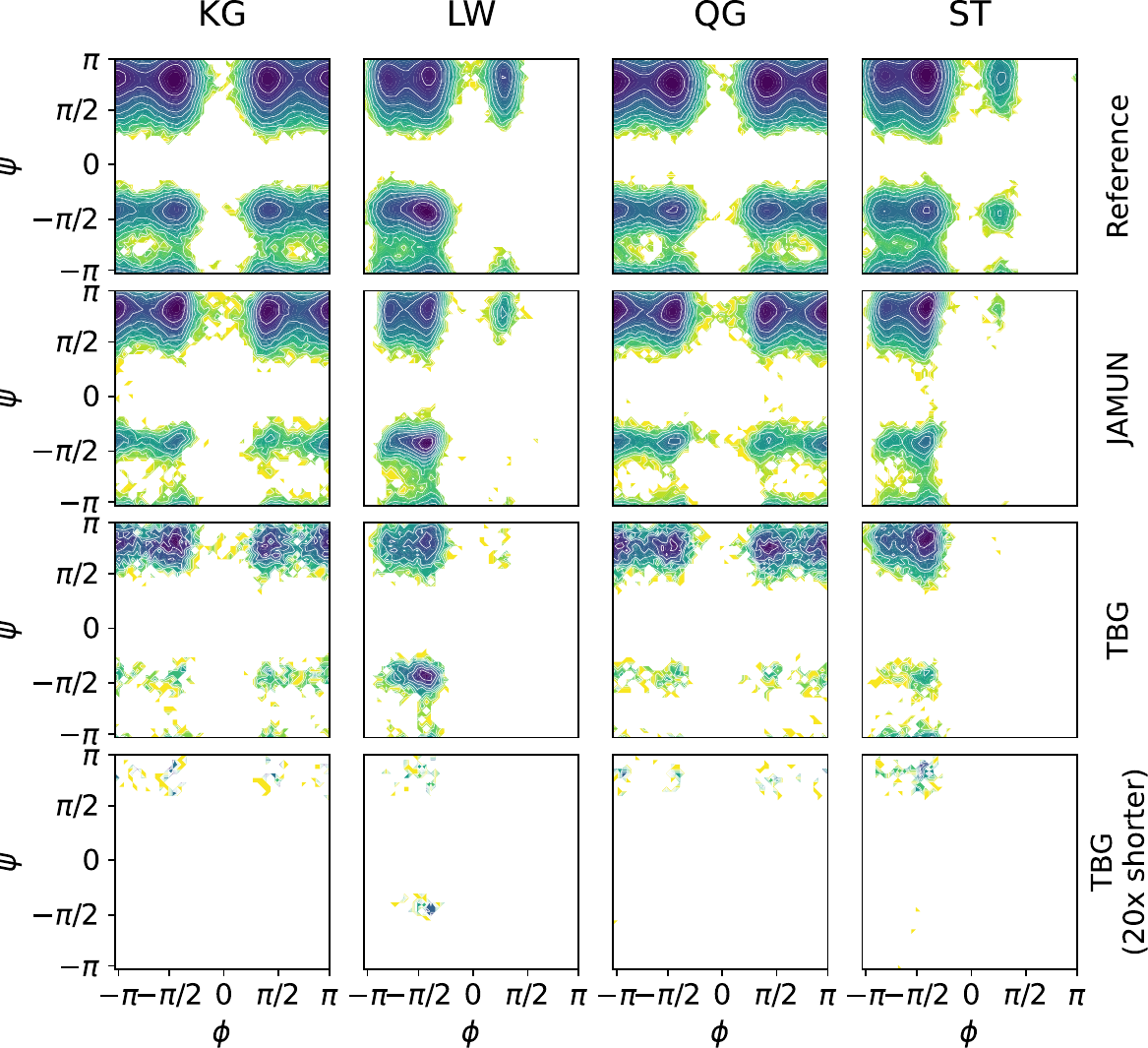}
    \caption{Ramachandran plots for $4$ randomly chosen test peptides on \dataset{Timewarp 2AA-Large}.}
    \label{fig:Timewarp_2AA_ramachandrans}
\end{figure}

\begin{figure}[h]
    \centering
    \includegraphics[width=0.9\linewidth]{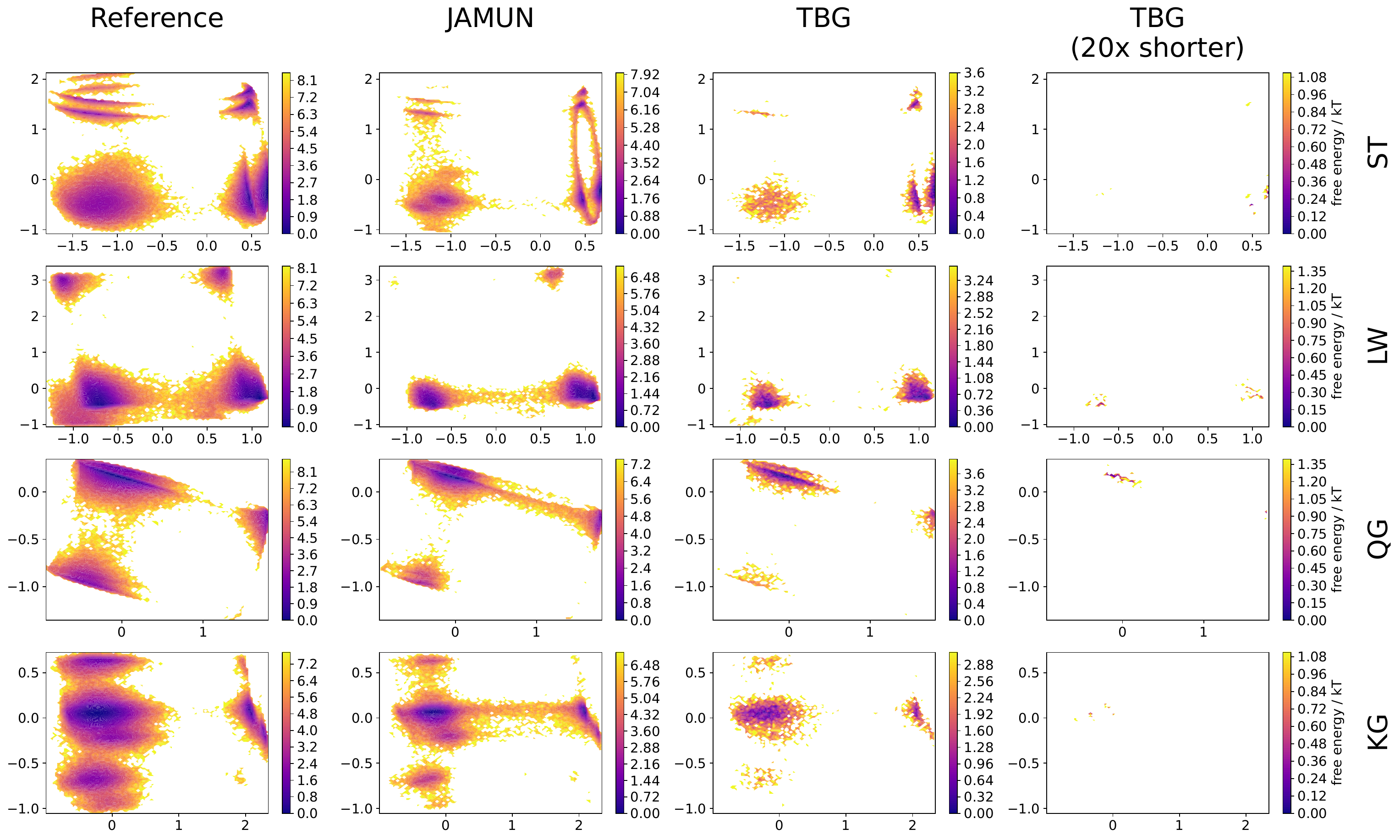}
    \caption{TICA-0,1 projections for $4$ randomly chosen test peptides on \dataset{Timewarp 2AA-Large}.}
    \label{fig:Timewarp_2AA_tica_projections}
\end{figure}

In \autoref{fig:Timewarp_2AA_ramachandrans} and \autoref{fig:Timewarp_2AA_tica_projections}, we visualize the TICA-0,1 projections and Ramachandran plots for randomly chosen test peptides from \dataset{Timewarp 2AA-Large}, highlighting that TBG misses certain basins that JAMUN is able to sample. 

\begin{figure}[h]
    \centering
    \includegraphics[width=0.9\linewidth]{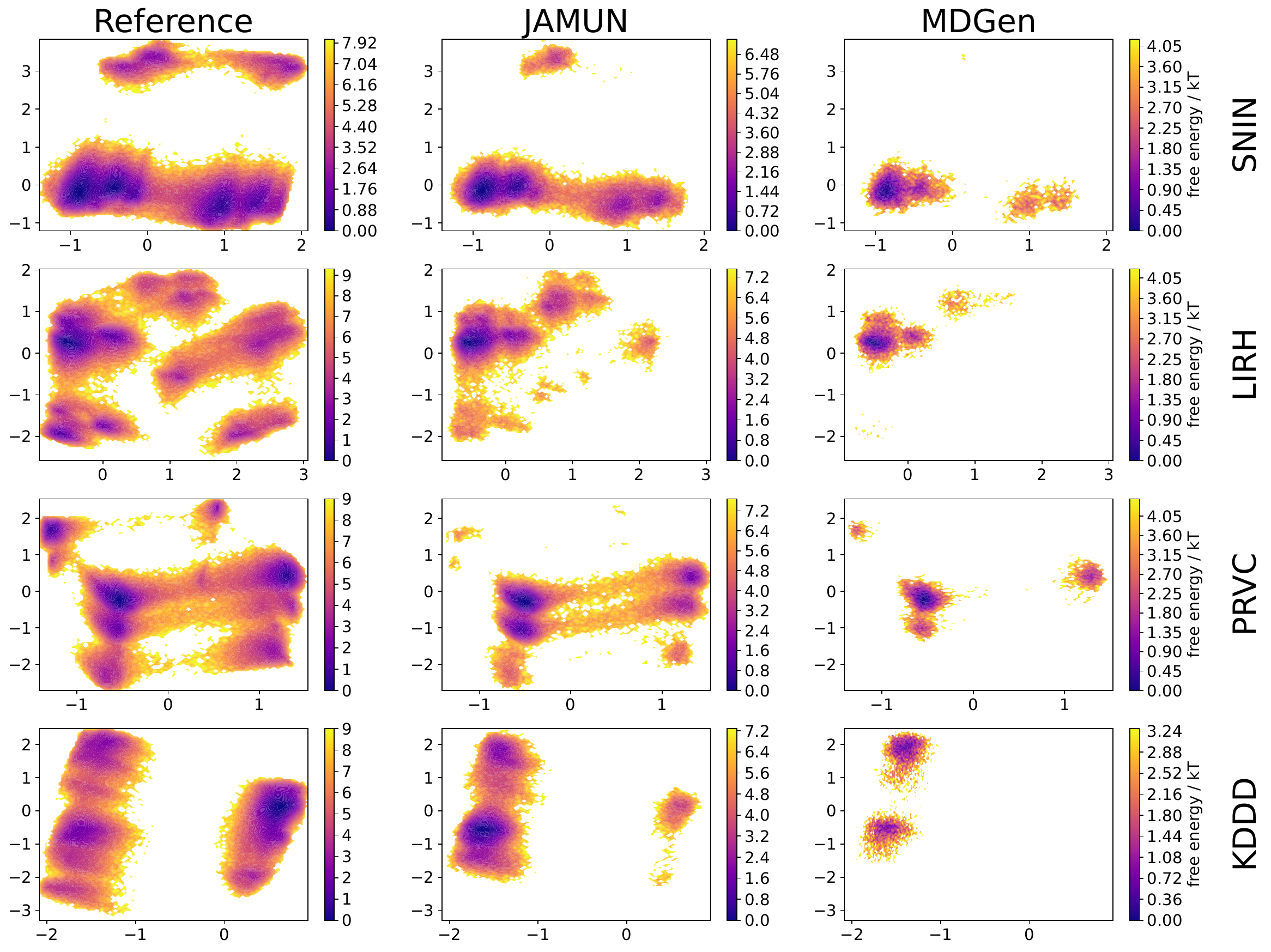}
    \caption{TICA-0,1 projections for $4$ randomly chosen test peptides from the \dataset{MDGen 4AA-Explicit} dataset.}
    \label{fig:MDGen_4AA_tica_projections}
\end{figure}

\begin{figure}[h]
    \centering
    \includegraphics[width=0.8\linewidth]{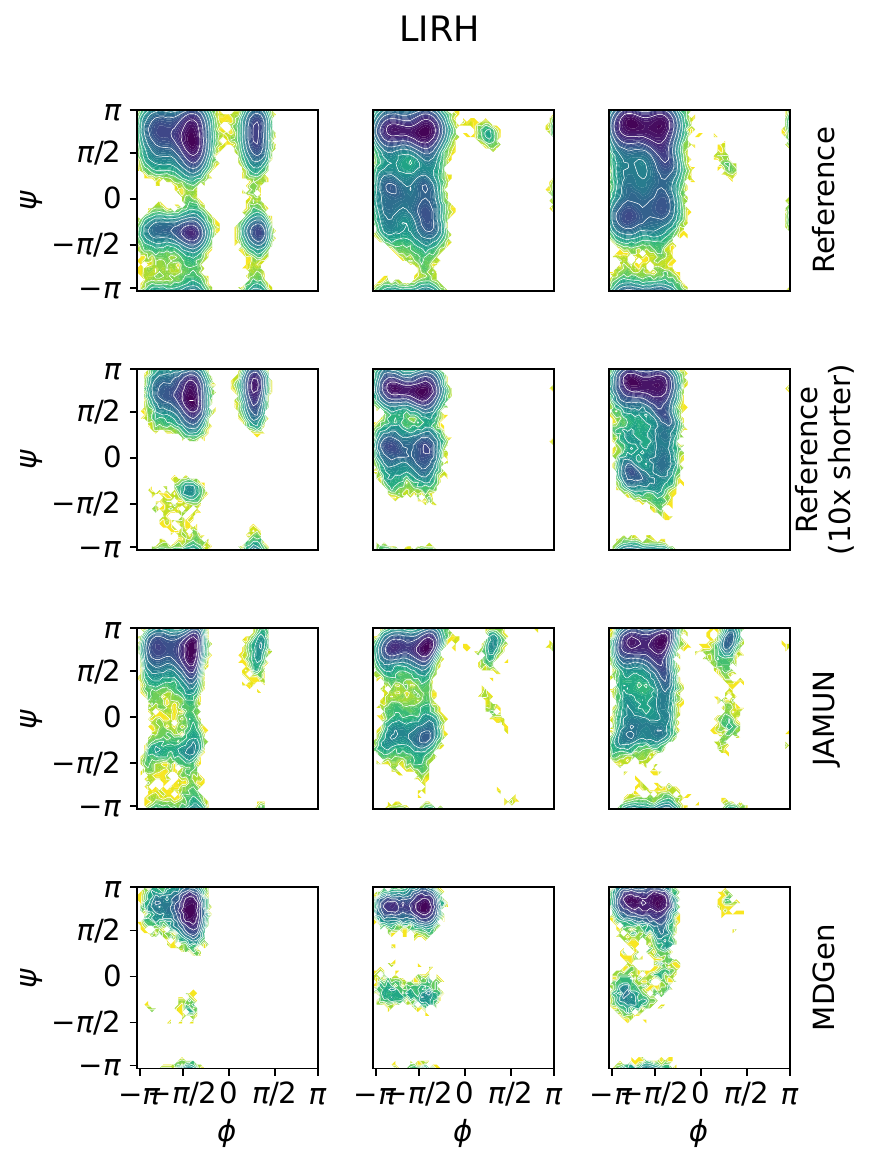}

    \caption{Ramachandran plots for JAMUN and MDGen on randomly chosen test peptide \texttt{LIRH} from the \dataset{MDGen 4AA-Explicit} dataset.}
    \label{fig:MDGen_4AA_ramachandrans}
\end{figure}

\autoref{tab:MDGen_4AA_jsd_table} shows that JAMUN is very competitive with MDGen on the JSD metrics. In fact, \autoref{fig:MDGen_4AA_tica_projections} shows that MDGen is missing some basins that JAMUN is able to sample. 
On the other hand, \autoref{fig:MDGen_4AA_ramachandrans} show an example where JAMUN hallucinates a basin.


\subsection{Assessing Generalization over Peptide Lengths with \dataset{Uncapped 5AA}}
\label{sec:uncapped-5AA-results}

\begin{figure}[h]
    \centering
    \includegraphics[width=\linewidth]{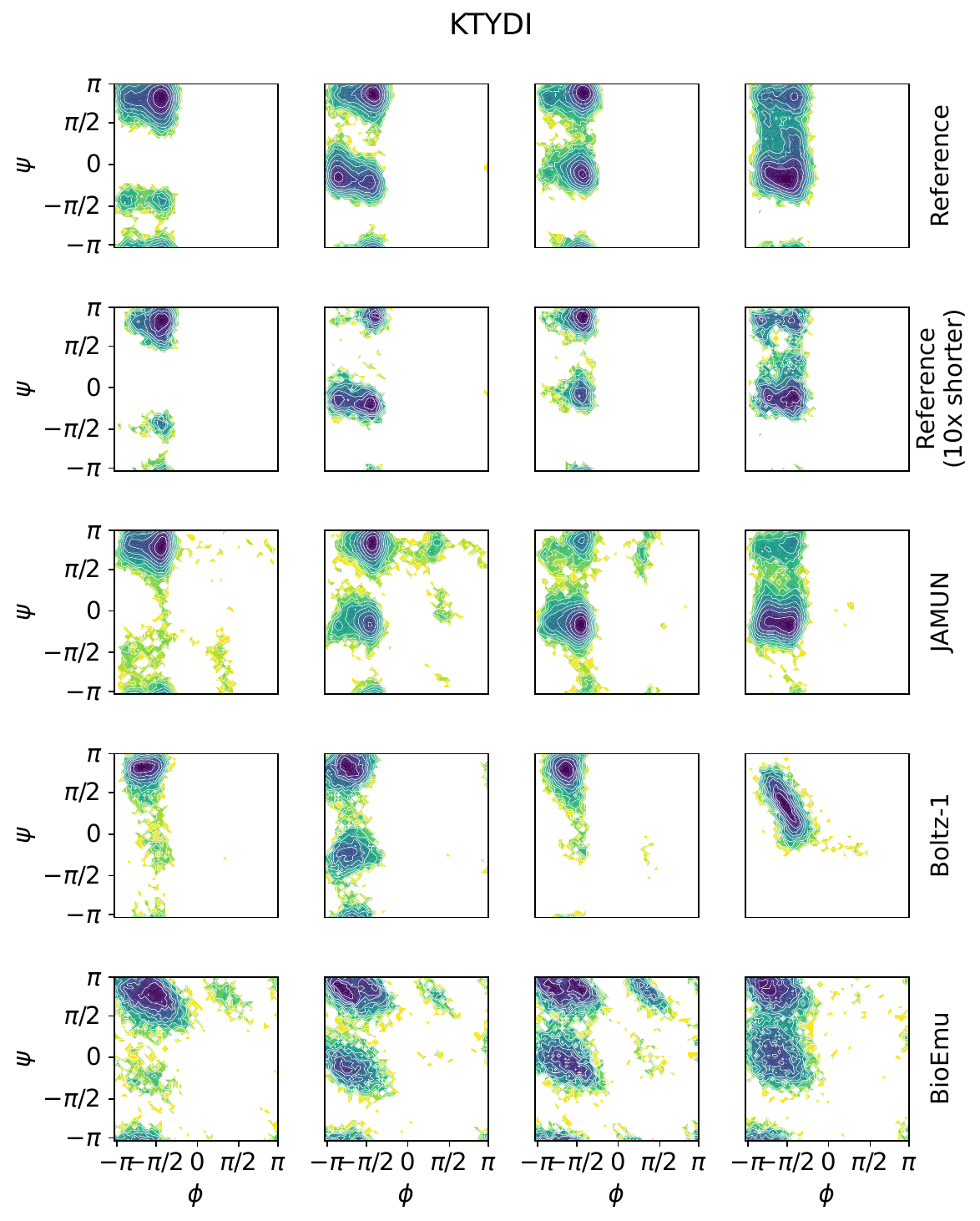}
    \caption{Ramachandran plots for test peptides \texttt{NRLCQ}, \texttt{VWSPF} and \texttt{KTYDI} in \dataset{Uncapped 5AA}.}
    \label{fig:Uncapped_5AA_ramachandrans}
\end{figure}

The message-passing architecture of JAMUN enables it to operate on molecules of larger sizes than it was originally trained on. Here, we test whether JAMUN can generalize to peptides of lengths beyond its training set. This is a challenging task, and one that we believe conformational generation models have not been adequately benchmarked on. 
Unfortunately, neither TBG nor MDGen transfer to \dataset{Uncapped 5AA} due to fixed-length absolute positional embeddings, despite our best efforts. In particular, we tried initializing the positional embeddings to support longer peptides, but this resulted in broken topologies in the resulting samples. Instead, we choose Boltz-1 and BioEmu which support sampling on \dataset{Uncapped 5AA} to compare against JAMUN trained on \dataset{Timewarp 4AA-Large}. 

\begin{figure}[h]
    \centering
    \includegraphics[width=0.9\linewidth]{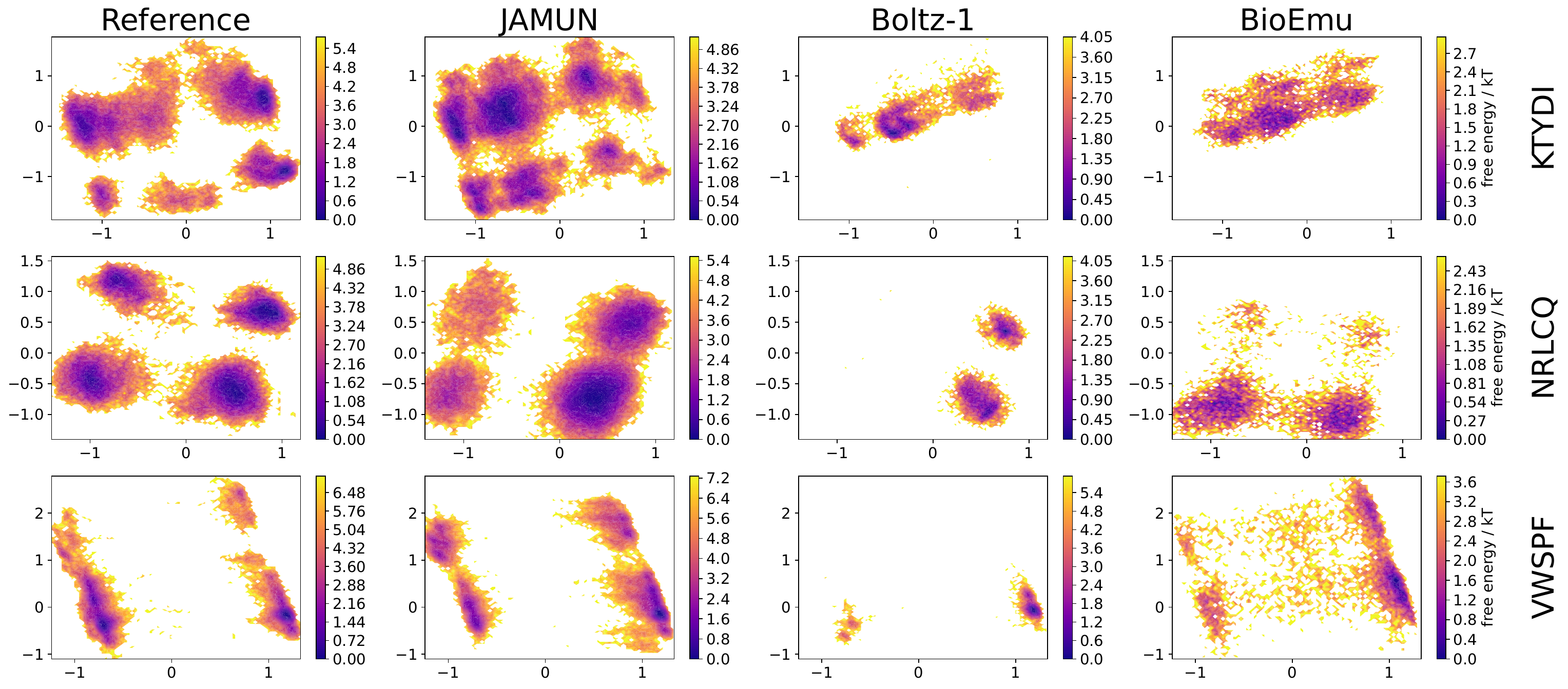}
    \caption{TICA-0,1 projections for three test peptides in \dataset{Uncapped 5AA}.}
    \label{fig:Uncapped_5AA_tica_projections}
\end{figure}

Surprisingly, we find that the JAMUN model \emph{trained only on 4AA peptides can accurately predict ensembles for 5AA peptides}.  \autoref{fig:Uncapped_5AA_ramachandrans} and \autoref{fig:Uncapped_5AA_tica_projections} show that JAMUN is able to recover most states and even reproduce relative probabilities. Interestingly, the same experiment does not work if we train on \dataset{Timewarp 2AA-Large} instead, suggesting that the 2AA reference MD data may not be informative enough to generalize from. We find that JAMUN is also not able to generalize well to much longer peptides (such as 10AA) due to the fact that the 4AA peptides that it was trained on lack any sort of secondary structure.

On the other hand, we find that Boltz-1 is unable to sample the diversity of peptide conformations. This is not entirely surprising as Boltz-1 was not trained on any MD data, as we noted before. Further, Boltz-1 also utilizes a common pair representation across all diffusion samples, as computed by its Pairformer stack. The pair representation intuitively represents the residue-wise distance matrix, and thus encodes a significant portion of the geometry. Keeping this representation fixed possibly prevents the sampling of large conformational changes.

Surprisingly, BioEmu also seems to struggle in this setting, even when considering distributions of backbone torsion angles only. This suggests that BioEmu cannot capture the relative flexibility of smaller peptides, even when trained on MD data for much larger proteins.

Quantitatively, \autoref{tab:Uncapped_5AA_jsd_table} shows that JAMUN significantly outperforms Boltz-1 and BioEmu in terms of the JSD metrics. Further, as seen in \autoref{tab:sampling-times}, JAMUN is roughly $5 \times$ faster than Boltz-1, and is roughly $60 \times$ faster than BioEmu when we additionally perform the side-chain reconstruction with H-Packer. 


\subsection{Generating Conformational Ensembles of Macrocyclic Peptides in \dataset{Cremp}}
\label{sec:macrocycles}

\begin{figure}[h!]
    \centering
    \includegraphics[width=0.7\linewidth]{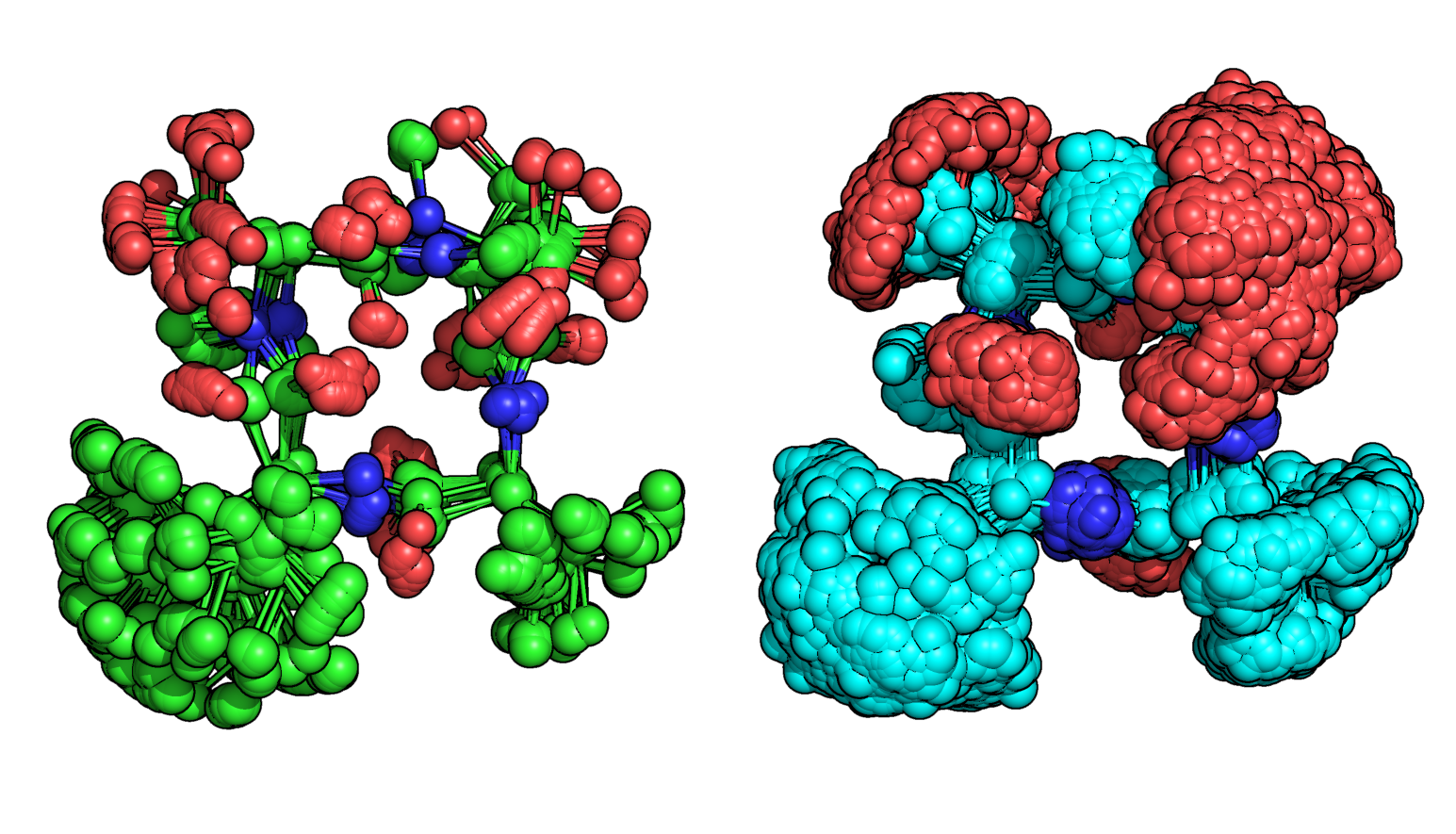}
    \caption{A 3D visualization of the CREMP (left, green backbone) and  JAMUN (right, cyan backbone) \texttt{MeS.MeS.V.L} macrocycle.}
    \label{fig:cremp_jamun_3d}
\end{figure}
    
\begin{figure}[h]
    \centering
    \includegraphics[width=0.7\linewidth]{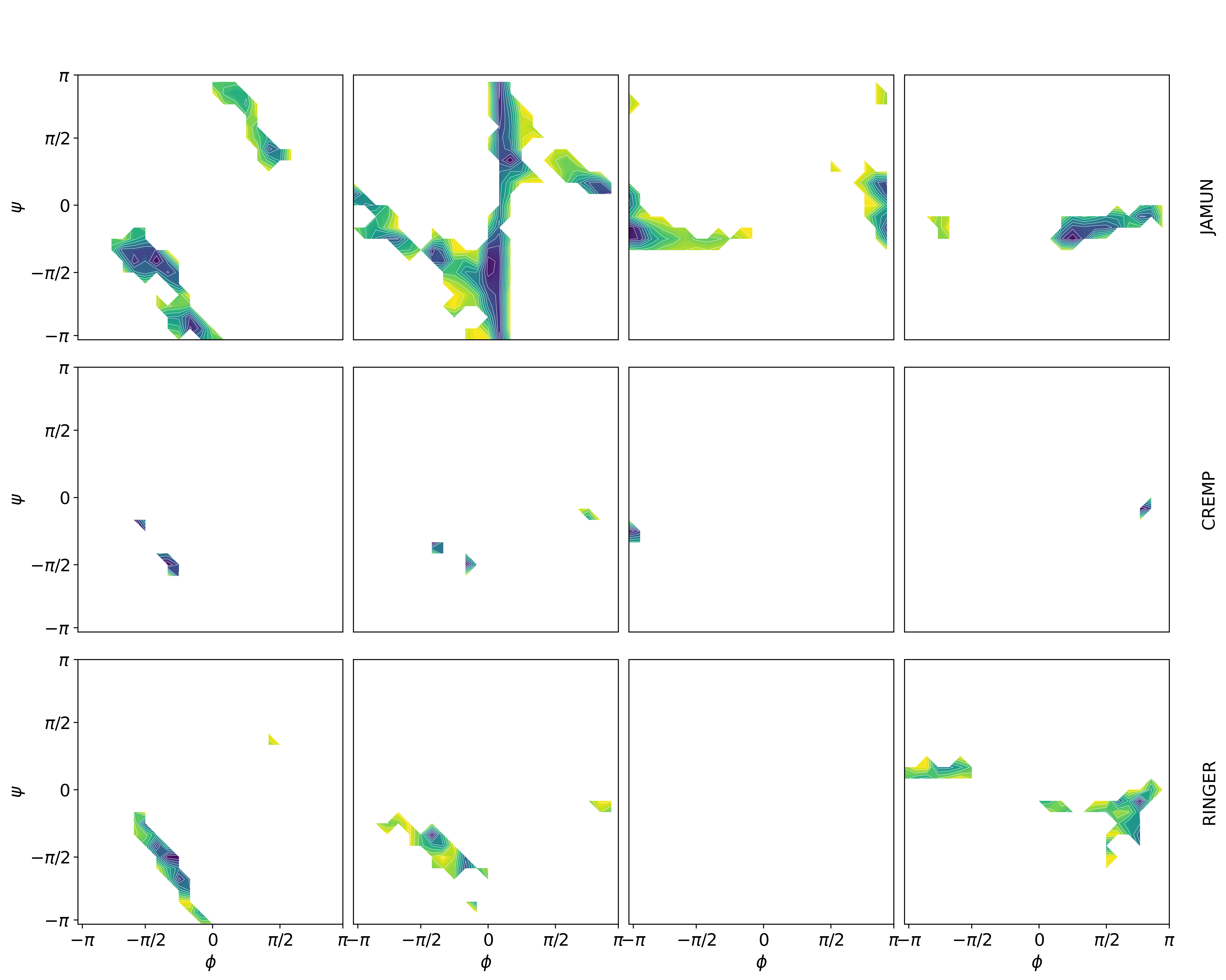}
        \label{fig:4mer_mcp}
    \caption{Ramachandran plots for JAMUN, CREMP, and RINGER samples of the 4AA \texttt{MeS.MeS.V.L} macrocycle.}
    \label{fig:4mer_macrocycle}
\end{figure}

\begin{figure}[h]
    \centering
    \begin{subfigure}[c]{0.7\linewidth}
        \centering
        \includegraphics[width=\linewidth]{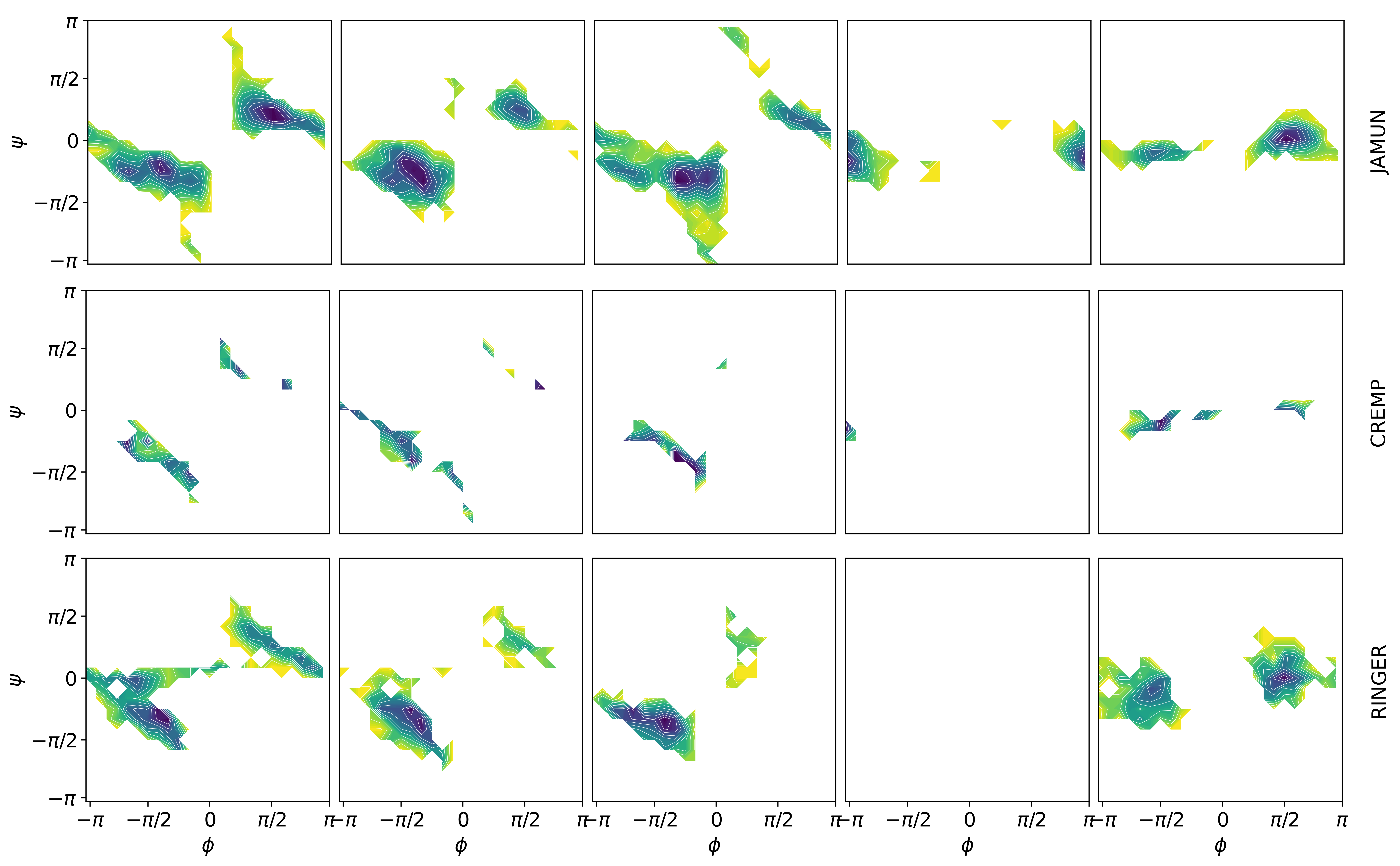}
        \label{fig:5mer_mcp}
    \end{subfigure}
\caption{Ramachandran plots for JAMUN, CREMP, and RINGER samples of the 5AA \texttt{F.Q.L.G.Met} macrocycle.}
\label{fig:longer_macrocycles}
\end{figure}

\begin{figure}[h]
    \centering
    \begin{subfigure}[c]{0.7\linewidth}
        \centering
        \includegraphics[width=\linewidth]{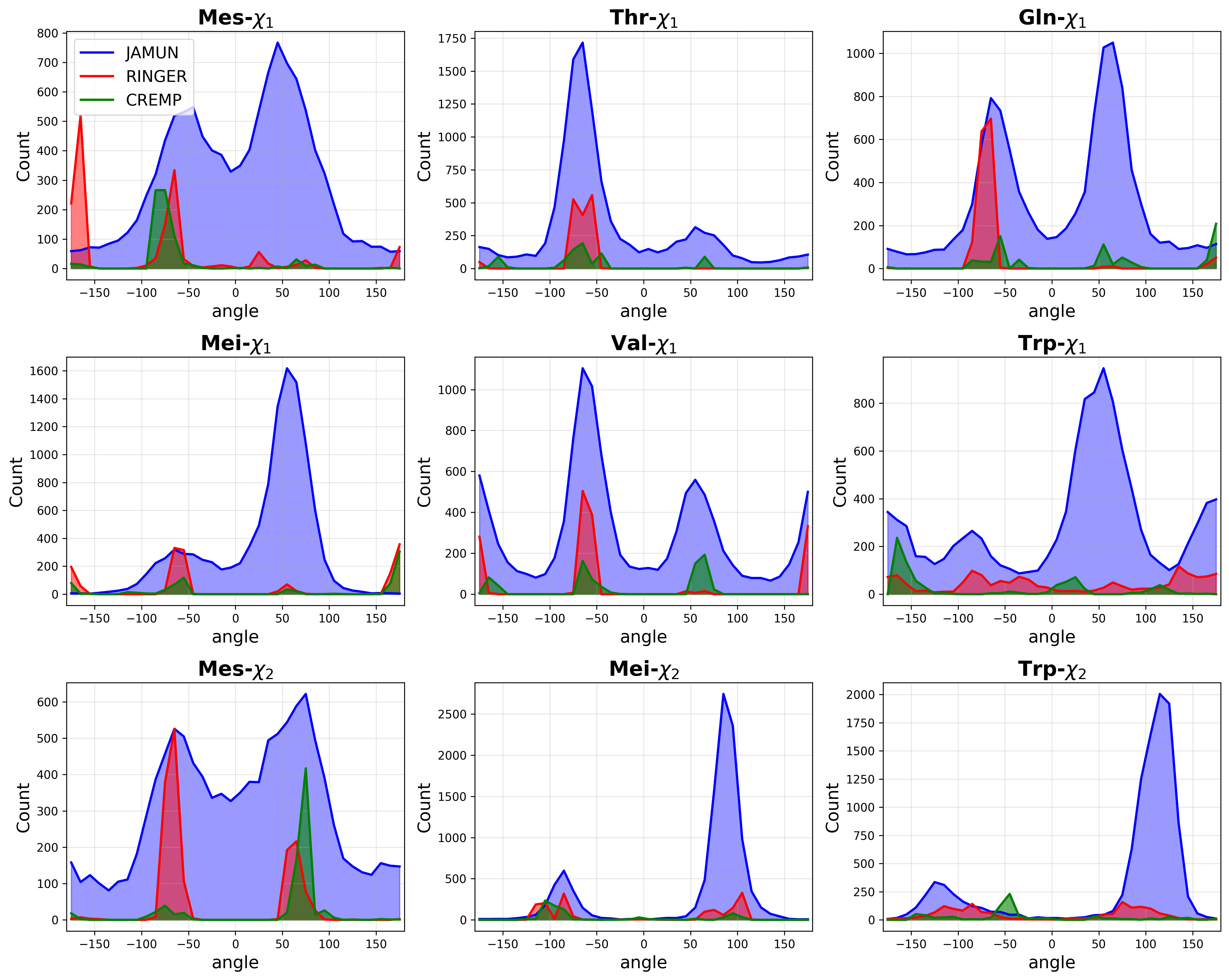}
        \label{fig:6mer_mcp_sidechains}
    \end{subfigure}\vspace{1em}
    \begin{subfigure}[c]{0.7\linewidth}
        \centering
        \includegraphics[width=\linewidth]{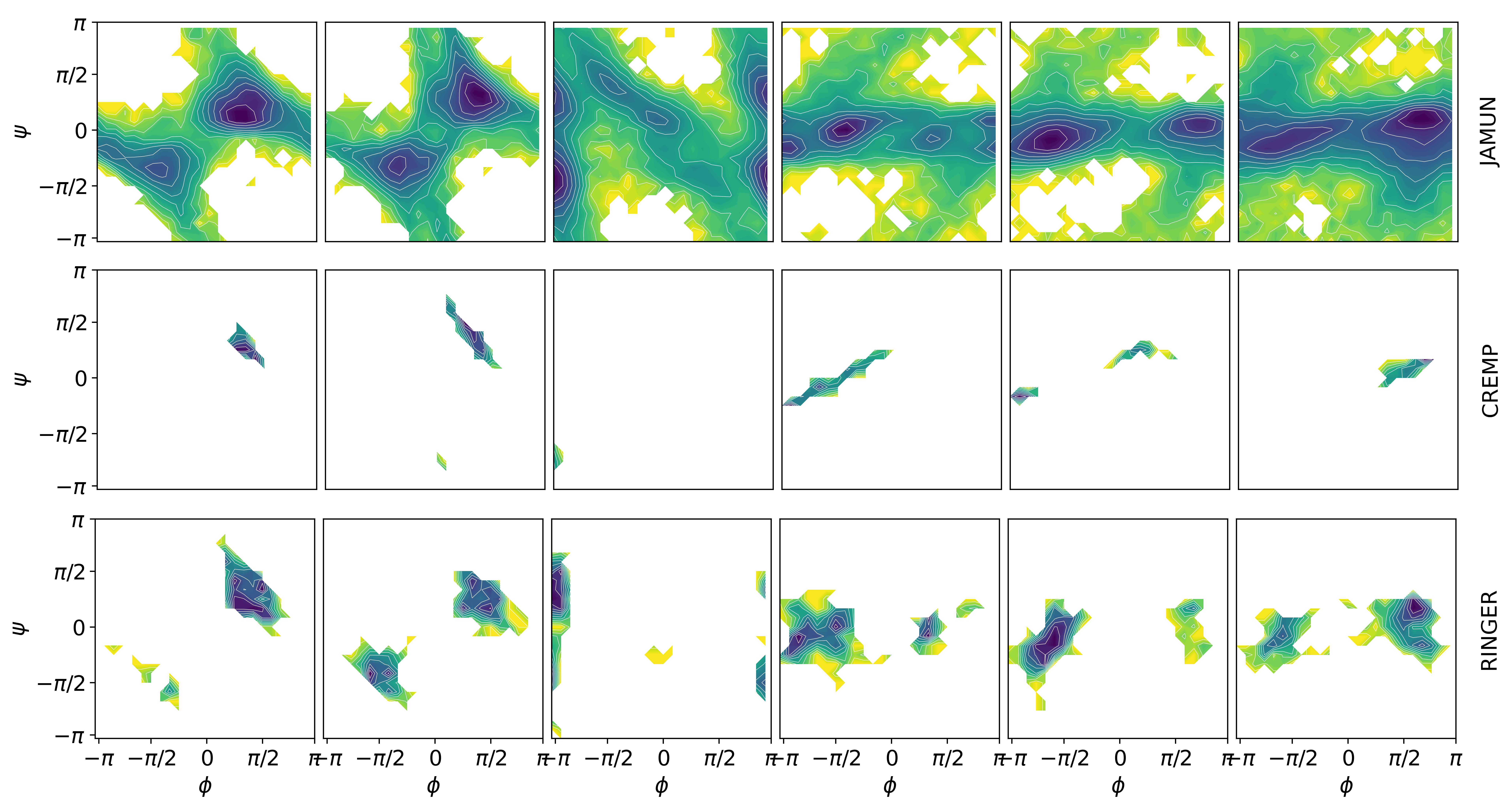}
        \label{fig:6mer_mcp}
    \end{subfigure}
\caption{Sidechain dihedral $\chi$ angles histogrammed for JAMUN, CREMP, and RINGER samples of the 6AA \texttt{Mes.T.Q.Mei.V.W} macrocycle.}
\label{fig:chiangles_macro}
\end{figure}


\todo{Add more discussion of RINGER here, including what changes Ameya made: decreasing batch size, increasing positional embedding length to support inference on 6AA. Highlight the cyclic attention as a reason why RINGER might generalize.}

\autoref{fig:4mer_macrocycle} and \autoref{fig:longer_macrocycles} illustrates the transferability of JAMUN to macrocyclic peptides.
It is clear that we are able to recover most of the sampled basins, even though it does seem that there are new basins uncovered. We note that our outputs look significantly more diffusive than the ground truth. The main reason for this is that the CREST data is filtered to only keep local minima, dropping the rest of the MD trajectory. Despite this, JAMUN is able to extrapolate and traverse the space between basins.
For a fair comparison, we retrained RINGER on \dataset{Cremp 4AA}: we reduced the batch size from $8192$ to $64$, and increased the maximum length of the positional embedding to $36$ to enable the sampling of longer peptides. We found that these changes significantly improved generalization.


We find that for \emph{unseen} 4AA macrocycles (\autoref{fig:4mer_macrocycle}), we are able to recover all the basins sampled by MD. Additionally, we ran inference on \dataset{Cremp 5AA} and \dataset{Cremp 6AA} with the same model to test length generalizability. In \autoref{fig:longer_macrocycles} and \autoref{fig:chiangles_macro}, we show the backbone and side-chain dihedral ($\chi$) angles of a randomly chosen 6AA macrocycle.  
We see that JAMUN recapitulates most modes while yielding a smoother distribution than RINGER. This is in spite of the fact that JAMUN is trained on all-atom configurations (including side-chain dihedrals), whereas RINGER relies on internal coordinates followed by a least-squares quadratic optimization for mapping to Cartesian coordinates. It is also interesting to note that both RINGER and JAMUN occasionally hallucinate the same modes. While this is a preliminary study, it points to the potential of JAMUN being useful even in a low-data regime.

\section{Conclusion}
\label{sec:conclusions}

We present JAMUN, a walk-jump sampling model for generating ensembles of molecular conformations, outperforming the state-of-the-art TBG model, and competitive with the performance of MDGen with no protein-specific parametrization. This represents an important step toward the ultimate goal of a transferable generative model for protein conformational ensembles.
Performing MD in the noised space gives the model a clear physics interpretation, and allows faster decorrelation (and hence, sampling) than classical MD.

The model has some limitations that motivate future work. While it is highly transferable in the space of two to six amino acid peptides, scaling up is likely to require more exploration and intensive data generation in future work. Additionally, while the current $SE(3)$-equivariant denoiser architecture works well, further development of the denoising network could speed up sampling. Alternative jump methods, such as multiple denoising steps (\`a la diffusion), could also serve to sharpen generation. Lastly, a promising direction that has not yet been explored is the application of classical enhanced sampling methods, such as metadynamics, but in the smoothed $\mathcal{Y}$ space.

\clearpage


\bibliography{main.bib}
\bibliographystyle{abbrvnat}

\newpage
\appendix
\onecolumn
\section{Comparison of Walk-Jump Sampling Against Full Diffusion}
\label{sec:diffusion-comparison}

\autoref{sec:experiments} contains a comparison of JAMUN with existing baselines, but does not directly address the comparison of the techinique of walk-jump sampling with standard diffusion. We perform this comparison here; we fix the JAMUN architecture and directly compare walk-jump sampling (at a single noise level) to full diffusion (which samples a range of noise levels from very large to very small).

We train a diffusion model with the same JAMUN architecture on \dataset{Timewarp 2AA}. We then sample from this model using both diffusion (specifically, the ODE sampler with a noise schedule of $64$ steps from $0.01\si{\angstrom}$ to $10\si{\angstrom}$ using the Heun second order method as recommended by EDM \citep{karras2022edm}) and walk-jump sampling. We compare the Jensen-Shannon divergence (JSD) of the backbone torsions averaged over the \dataset{Timewarp 2AA} test set as a function of number of samples and number of function evaluations (NFE):

\begin{figure}[h]
    \centering
    \includegraphics[width=0.49\linewidth]{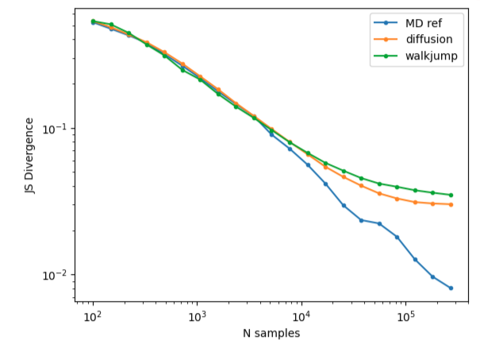}
    \includegraphics[width=0.49\linewidth]{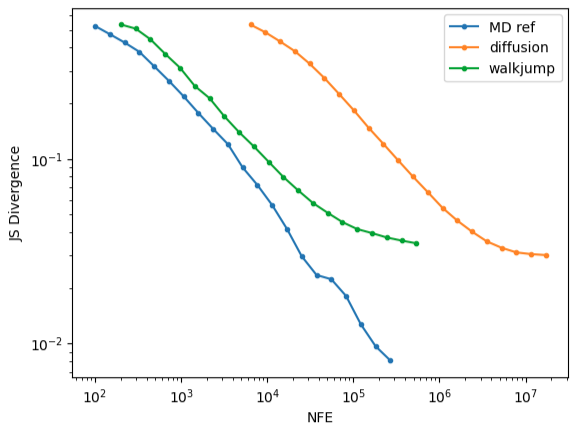}
    \caption{Comparison between diffusion, walk-jump and the ground truth MD in terms of JS divergence of backbone torsions to the full ground truth MD distributions, as a function of (left) number of samples and (right) number of function evaluations.}
    \label{fig:diffusion-nsamples}
\end{figure}

\begin{table*}[h]
\centering
\begin{tabular}{cccc}
\toprule
Sampler & Number of Samples & NFE & JSD-Backbone Torsions $\downarrow$ \\
\midrule
Walk-Jump & 3149 & 6298 & 0.1501 \\
Diffusion & 3149 & 399923 & 0.1363 \\
Walk-Jump & 200000 & 400000 & 0.0496 \\
Diffusion & 200000 & 25400000 & 0.0460 \\
\bottomrule
\end{tabular}
\caption{Comparison of walk-jump and diffusion sampling with the same model. Diffusion obtains a better JSD metric than walk-jump sampling, but at the cost of $\approx 30$ more sampling time.}
\label{tab:sampler_comparison}
\end{table*}

\autoref{tab:sampler_comparison} contains a summary of the key comparison.
Essentially, we find that walk-jump sampling is faster with only minor loss in fidelity; because it works in a partially noised space, instead of having to generate every sample from an uninformative Gaussian prior over many steps.

\section{Physical Validity and Energy Analysis}
\label{sec:physical-validity}

 Here, we perform a physical validity analysis on the generated JAMUN samples, using the popular Posebusters \citep{posebusters} package. We randomly selected $20$ unseen test peptides from the \dataset{MDGen 4AA-Explicit} dataset for this analysis.

\begin{table*}[h]
\centering
\begin{tabular}{cc}
\toprule
Posebusters Metric & Average Pass Rate \\
\midrule
Valid Bond Lengths & 97.0\% \\
Valid Bond Angles & 99.4\% \\
Internal Steric Clash & 100.0\% \\
Internal Energy & 97.5\% \\
Overall & 94.7\% \\
\bottomrule
\end{tabular}
\caption{Average pass rates for Posebusters metrics on JAMUN samples.}
\label{tab:posebusters_avg}
\end{table*}

We see that the bond lengths are correctly captured with high probability by JAMUN. Furthermore, the overall quality of the JAMUN samples is high. For a finer-grained view into the performance across the 20 test peptides, we report the empirical CDFs of the pass rates:

{
\renewcommand{\arraystretch}{1.5}
\begin{table*}[h]
\centering
\begin{tabular}{ccccccc}
\toprule
Posebusters Pass Rate & $>90\%$ & $>92\%$ & $>94\%$ & $>96\%$ & $>98\%$ & $100\%$ \\
\midrule
Valid Bond Lengths & $\frac{20}{20}$ & $\frac{18}{20}$ & $\frac{17}{20}$ & $\frac{14}{20}$ & $\frac{10}{20}$ & $\frac{5}{20}$ \\
Valid Bond Angles & $\frac{20}{20}$ & $\frac{20}{20}$ & $\frac{20}{20}$ & $\frac{20}{20}$ & $\frac{18}{20}$ & $\frac{14}{20}$ \\
Internal Steric Clash & $\frac{20}{20}$ & $\frac{20}{20}$ & $\frac{20}{20}$ & $\frac{20}{20}$ & $\frac{20}{20}$ & $\frac{20}{20}$ \\
Internal Energy & $\frac{20}{20}$ & $\frac{19}{20}$ & $\frac{19}{20}$ & $\frac{17}{20}$ & $\frac{10}{20}$ & $\frac{5}{20}$ \\
Overall & $\frac{19}{20}$ & $\frac{14}{20}$ & $\frac{13}{20}$ & $\frac{11}{20}$ & $\frac{6}{20}$ & $\frac{2}{20}$ \\
\bottomrule
\end{tabular}
\caption{Empirical CDF of pass rates across 20 test peptides.}
\label{tab:posebusters_cdf}
\end{table*}
}

Next, we compute the force field energies for JAMUN samples. We add hydrogen atoms using OpenMM's PDBFixer \citep{Openmm_1}, and compute energies using the \texttt{amber14} force field. We find that the energies of the JAMUN samples overlap well with those of the reference MD samples:
\begin{table*}[h]
\centering
\begin{tabular}{ccc}
\toprule
Sequence & \dataset{MDGen 4AA-Explicit} & JAMUN \\
\midrule
\texttt{FHSE} & $-675.6 \pm 20.1$ & $-633.2 \pm 113.3$ \\
\texttt{FKKL} & $-699.0 \pm 24.0$ & $-562.5 \pm 192.1$ \\
\texttt{FLRH} & $-1272.7 \pm 18.9$ & $-1198.5 \pm 129.3$ \\
\texttt{FSDP} & $-697.5 \pm 23.7$ & $-687.6 \pm 91.5$ \\
\texttt{FSRK} & $-1333.0 \pm 21.9$ & $-1349.9 \pm 0.0$ \\
\texttt{GCIC} & $-557.5 \pm 21.4$ & $-538.8 \pm 56.4$ \\
\texttt{GGHN} & $-905.3 \pm 21.9$ & $-821.3 \pm 129.6$ \\
\texttt{GLIL} & $-743.3 \pm 20.7$ & $-711.6 \pm 71.8$ \\
\texttt{HELI} & $-794.2 \pm 25.5$ & $-780.6 \pm 73.5$ \\
\texttt{HENV} & $-1156.9 \pm 17.3$ & $-1123.4 \pm 148.7$ \\
\texttt{HTIQ} & $-762.9 \pm 17.0$ & $-726.8 \pm 105.8$ \\
\texttt{IAMI} & $-428.0 \pm 15.3$ & $-426.8 \pm 74.8$ \\
\texttt{IDRH} & $-1416.8 \pm 18.1$ & $-722.8 \pm 2608.1$ \\
\texttt{IHNV} & $-845.4 \pm 21.1$ & $-864.4 \pm 48.8$ \\
\texttt{IMRY} & $-1230.7 \pm 23.6$ & $-1100.6 \pm 207.0$ \\
\texttt{INVH} & $-793.6 \pm 22.9$ & $-745.3 \pm 128.6$ \\
\texttt{IPGD} & $-611.5 \pm 15.0$ & $-582.5 \pm 55.2$ \\
\bottomrule
\end{tabular}
\caption{Force field energies (in (\si{\kilo\joule\per\mole}) comparison between samples from \dataset{MDGen 4AA-Explicit} and JAMUN.}
\label{tab:energy_comparison}
\end{table*}
\section{Overview of Denoiser}
\label{sec:denoiser-details}

\begin{figure}[p]
    \centering
    \includegraphics[width=\linewidth]{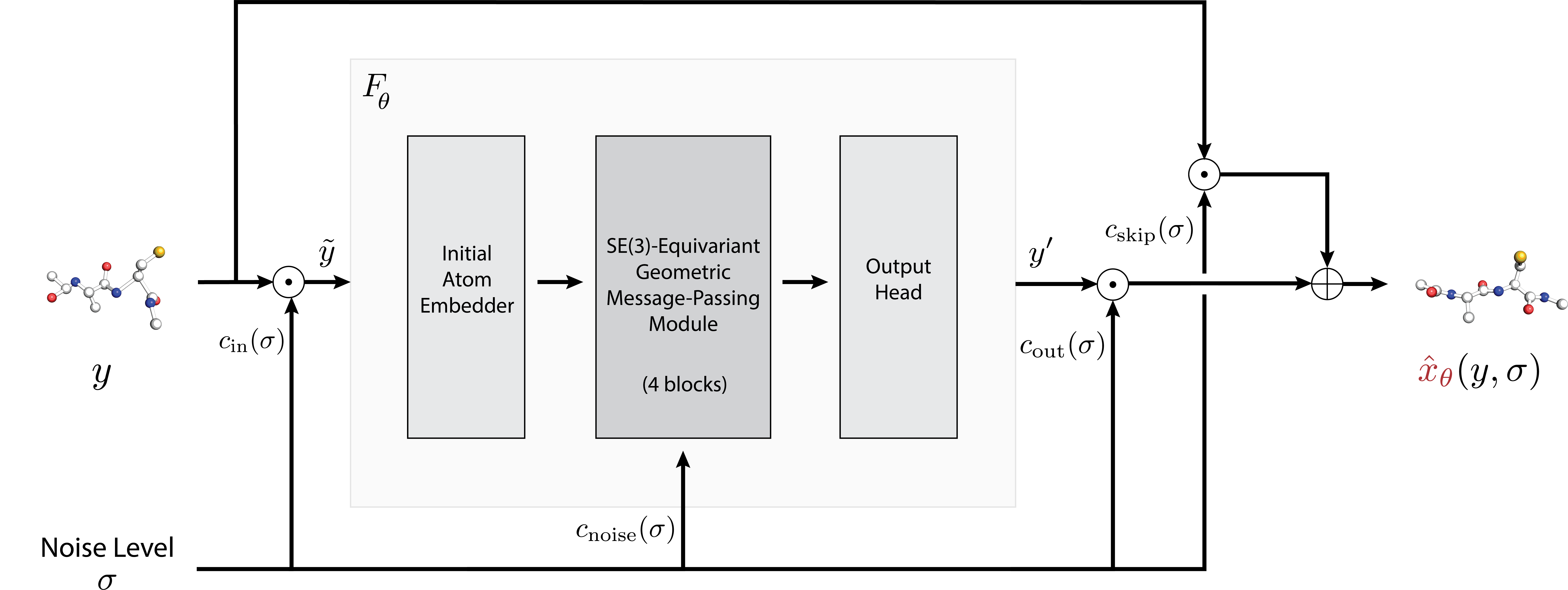}
    \caption{Overview of the denoiser network $\highlight[1]{\Dt}$. The submodule $\Ft$ sees input atom coordinates $\tilde{y} = \cin y$ and outputs predicted atom coordinates $y'$, which gets scaled and added to a noise-conditional skip connection to finally obtain $\highlight[1]{\Dt}(y)$.}
    \label{fig:denoiser-overview}
\end{figure}

\begin{figure}[h]
    \centering
    \includegraphics[width=0.88\linewidth]{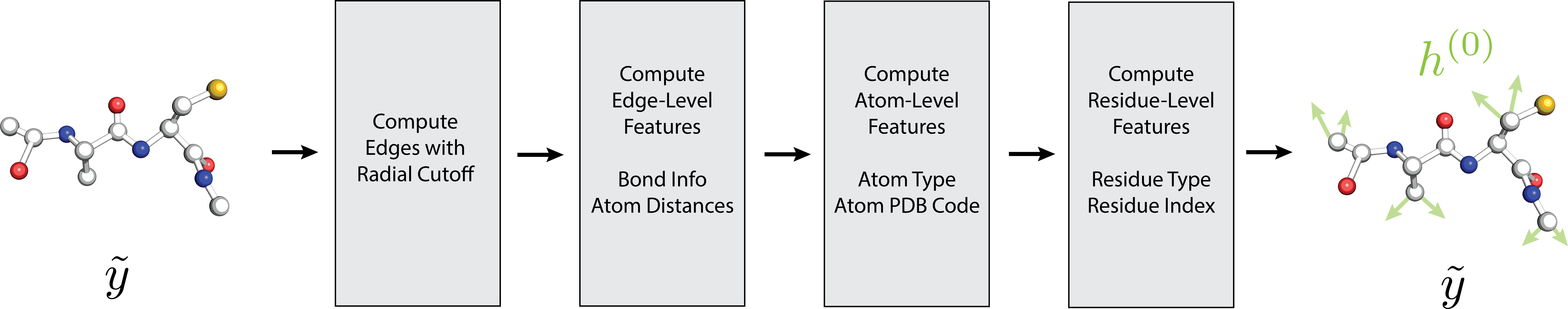}
    \caption{Overview of the initial embedder in the denoiser network, creating initial features $h^{(0)}$ at each atom and edge.}
    \label{fig:denoiser-initial}
\end{figure}

\begin{figure}[h]
    \centering
    \includegraphics[width=0.88\linewidth]{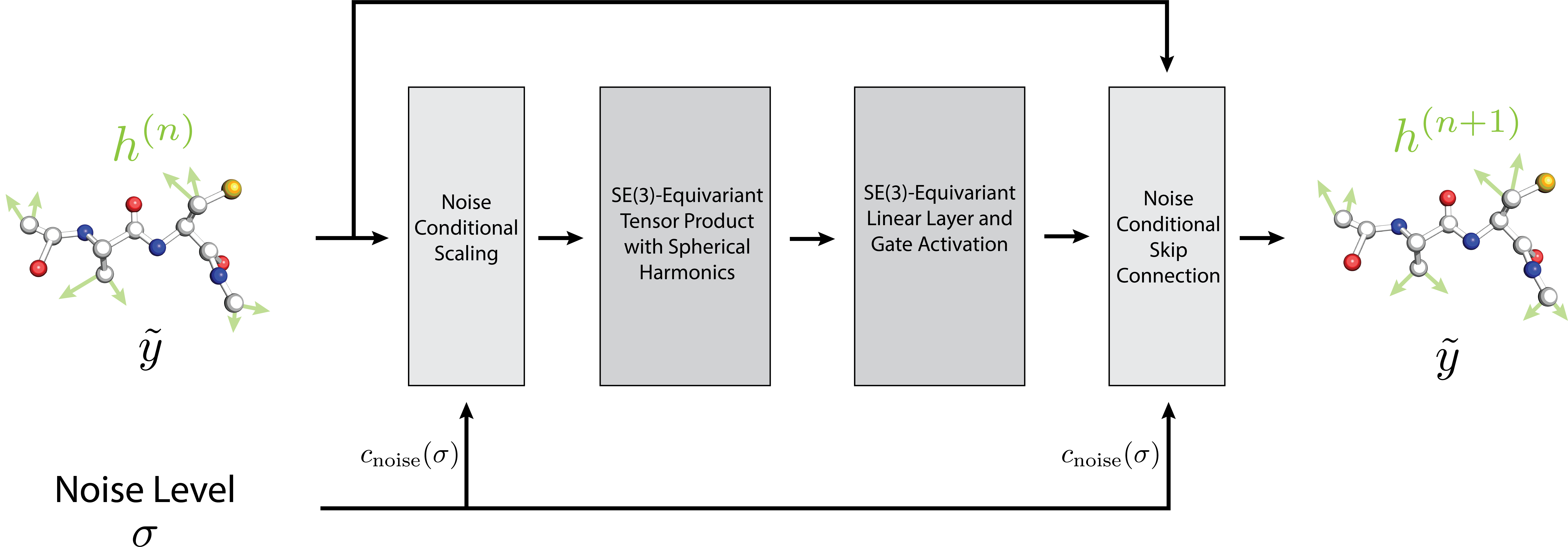}
    \caption{Overview of a single $SE(3)$-equivariant message-passing block (indexed by $n$) in the denoiser network. There are four such blocks iteratively updating the atom features from $h^{(0)}$ to $h^{(4)}$. The atom coordinates  denoted by $\tilde{y} = \cin y$ (and hence, the edge features) are unchanged throughout these blocks.}
    \label{fig:denoiser-message-passing}
\end{figure}

\begin{figure}[h]
    \centering
    \includegraphics[width=0.88\linewidth]{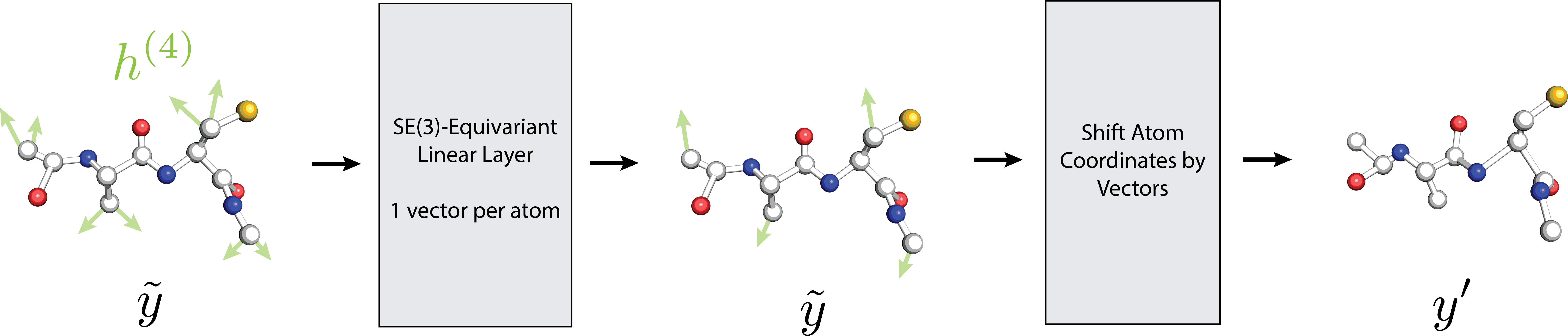}
    \caption{Overview of the output head, which predicts the coordinates $y' = \Ft(\cin y, \cnoise)$.}
    \label{fig:denoiser-output-head}
\end{figure}

The denoiser is a $SE(3)$-equivariant graph neural network, similar to NequIP \citep{nequip}.
The graph is defined by a radial cutoff of $10 \si{\angstrom}$ between positions in $y$. The overall computation performed by the denoiser is shown in \autoref{fig:denoiser-overview}. The initial embedding, message-passing and output head blocks are shown in \autoref{fig:denoiser-initial}, \autoref{fig:denoiser-message-passing} and \autoref{fig:denoiser-output-head} respectively.

The hidden features $h^{(n)}$ for $n = 0, \ldots, 4$ contain $120$ scalar and $32$ vector features per atom. We use spherical harmonics up to $l = 1$ for the tensor product across each edge. If $\otimes$ represents the tensor product, then each layer performs the following operation at each node $i$:
\begin{align}
    h_i^{(t + 1)} = \frac{1}{|\mathcal{N}_i|}\sum_{j \in \mathcal{N}_i} \operatorname{MLP}(\operatorname{GaussianEmbed}(\norm{\tilde{y}_j - \tilde{y}_i}), \operatorname{BondEmbed}((i, j) \text{ are bonded})\otimes \left(h_j^{(t)} \otimes Y^l\left(\frac{\tilde{y}_j - \tilde{y}_i}{\norm{\tilde{y}_j - \tilde{y}_i}}\right)\right)
\end{align}

For all datasets, we train and sample with $\sigma = 0.4$ \si{\angstrom}. 
For the Langevin dynamics (\autoref{eqn:baoab-update}), we set $M = 1$, friction of $\gamma = 1.0$ and a step size of $\Delta t = \sigma$.

We use the Adam optimizer with learning rate $.002$.  Models are trained with a batch size of $32$ over $2$ NVIDIA RTX A100 GPUs.

\clearpage
\section{Normalization}
\label{sec:normalization}
While our model is trained on a fixed noise level, we discuss the normalization for an arbitrary noise level $\sigma$. 

As the noise level $\sigma$ is increased, $y = x + \sigma \noise$ where $\noise \sim \mathcal{N}(0, \mathbb{I}_{N \times 3})$ expands in space.
Let $\tilde{y}$ represent the `normalized' input $y$, as seen by the network $\Ft$:
\begin{align}
    \tilde{y} = \cin  y
\end{align}
To control the expansion of $y$, $\cin$ is chosen such that the following property holds:
\begin{align}
    \label{eqn:input_norm}
    \EE_{\substack{(i, j) \sim \text{Uniform}(E) \\ \noise \sim \mathcal{N}(0, \mathbb{I}_{N \times 3})}}[\norm{\tilde{y}_i - \tilde{y}_j}^2] = 1
    \ \text{at all noise levels} \ \sigma.
\end{align}
Note that we define $E$ with a radial cutoff of $10 \si{\angstrom}$ over the noisy positions $y$:
\begin{align}
    \label{eqn:E}
    E = \{(i, j): \norm{{y}_i - {y}_j} \leq 10 \si{\angstrom}\}
\end{align}
(In practice, for the noise levels we choose here, the effect of adding noise on the radial cutoff is negligible.)

Our normalization is distinct from the normalization chosen by \citep{karras2022edm,karras2024edm2}, which normalizes $\norm{y}$ instead of the pairwise distances $\norm{{y}_i - {y}_j}$ as we do. The intuition behind this normalization is that the GNN model $\Ft$ does not operate on atom positions $y$ directly, but instead uses the relative vectors $y_i - y_j$ to account for translation invariance, and controlling this object directly ensures that the topology of the graph does not change with varying noise level $\sigma$.

To achieve this, we compute the input normalization factor as:
\begin{align}
    \cin = \frac{1}{\sqrt{C + 6\sigma^2}}
\end{align}
where $C = \EE_{(i, j) \sim \text{Uniform}(E)} \norm{x_i - x_j}^2$ can be easily estimated from the true data distribution. The full derivation can be found in \autoref{sec:Input Normalization}.

As the input is now appropriately normalized, the target output of the network $\Ft$ should also be appropriately normalized. A full derivation, found in \autoref{sec:Output Normalization}, leads to:
\begin{align}
    \cskip &=  \frac{C}{C + 6 \sigma^2}
\end{align}
\begin{align}
    \cout &= \sqrt{\frac{C \cdot 6 \sigma^2}{C + 6 \sigma^2}}
\end{align}

The noise normalization is chosen as identical to \citet{karras2022edm,karras2024edm2}:
\begin{align}
    \cnoise = \log_{10} \sigma
\end{align}

\subsection{Input Normalization}
\label{sec:Input Normalization}
Fix an $(i, j) \in E$ from \autoref{eqn:E}.
As $\noise_i, \noise_j \overset{\mathrm{iid}}{\sim} \mathcal{N}(0, \mathbb{I}_{3})$, we have $\noise_i - \noise_j \sim \mathcal{N}(0, 2\mathbb{I}_{3})$ from the closure of the multivariate Gaussian under linear combinations.
Thus, for each component $d$, we have: $(\noise_i - \noise_j)_{(d)} \sim \mathcal{N}(0, 2)$ and hence:
\begin{align}
    \EE_{\noise \sim \mathcal{N}(0, \mathbb{I}_{N \times 3})}[(x_i - x_j)^T (\noise_i - \noise_j)] &= \sum_{d = 1}^3 (x_i - x_j)_{(d)}\EE[(\noise_i - \noise_j)_{(d)}] = 0
    \\
    \EE_{\noise \sim \mathcal{N}(0, \mathbb{I}_{N \times 3})}[\norm{\noise_i - \noise_j}^2] &= \sum_{d = 1}^3 \EE[(\noise_i - \noise_j)_{(d)}^2] = 6
\end{align}
We can now compute:
\begin{align*}
    &\EE_{z}[\norm{\tilde{y}_i - \tilde{y}_j}^2] \\
    &= \cin^2  \EE_\noise[\norm{y_i - y_j}^2] \\
    &= \cin^2  \EE_\noise[\norm{x_i - x_j + \sigma (\noise_i - \noise_j)}^2] \\
    &= \cin^2  \left(\norm{x_i - x_j}^2 + 2 \sigma \EE_\noise[(x_i - x_j)^T (\noise_i - \noise_j)]
    + \sigma^2 \EE_\noise[\norm{\noise_i - \noise_j}^2] \right)\\
    &= \cin^2  \left(\norm{x_i - x_j}^2 + \sigma^2 \EE_\noise[\norm{\noise_i - \noise_j}^2] \right) \\
    &= \cin^2  \left(\norm{x_i - x_j}^2 + 6\sigma^2 \right).
    \numberthis
\end{align*}
Now, taking the expectation over all $(i, j) \in E$ uniformly:
\begin{align*}
    \EE_{\substack{(i, j) \sim \text{Uniform}(E) \\ \noise \sim \mathcal{N}(0, \mathbb{I}_{N \times 3})}}[\norm{\tilde{y}_i - \tilde{y}_j}^2] &= 
    \EE_{(i, j) \sim \text{Uniform}(E)}[\EE_\noise[\norm{\tilde{y}_i - \tilde{y}_j}^2]] \\
    &= \cin^2  \left(\EE_{(i, j) \sim \text{Uniform}(E)} \norm{x_i - x_j}^2 + 6\sigma^2 \right)
    \numberthis
    \label{eqn:cin_orig}
\end{align*}
Let $C = \EE_{(i, j) \sim \text{Uniform}(E)} \norm{x_i - x_j}^2$, which we estimate from the true data distribution. Then, from \autoref{eqn:cin_orig} and our intended normalization given by \autoref{eqn:input_norm}:
\begin{align}
    \cin = \frac{1}{\sqrt{C + 6\sigma^2}}
\end{align}

\subsection{Output Normalization}
\label{sec:Output Normalization}
The derivation here is identical to that of \citep{karras2022edm,karras2024edm2}, but with our normalization.
The denoising loss at a single noise level is:
\begin{align}
    \label{eqn:denoising_loss_single}
    \mathcal{L}(\Dt, \sigma) = \EE_{X \sim p_X} \EE_{\noise \sim \mathcal{N}(0, \mathbb{I}_{N \times 3})} [\norm{\Dt(X + \sigma \noise, \sigma) - X}^2]
\end{align}
which gets weighted across a distribution $p_\sigma$ of noise levels by (unnormalized) weights $\lambda(\sigma)$:
\begin{align*}
    \mathcal{L}(\Dt) &= \EE_{\sigma \sim p_\sigma} [\lambda(\sigma)\mathcal{L}(\Dt, \sigma)] 
    \\
    &= \EE_{\sigma \sim p_\sigma} \EE_{X \sim p_X} \EE_{\noise \sim \mathcal{N}(0, \mathbb{I}_{N \times 3})} [\lambda(\sigma) \norm{\Dt(X + \sigma \noise, \sigma) - X}^2]
    \\
    &= \EE_{\sigma \sim p_\sigma} \EE_{X \sim p_X} \EE_{Y \sim \mathcal{N}(X, \sigma^2\mathbb{I}_{N \times 3})} [\lambda(\sigma) \norm{\Dt(Y, \sigma) - X}^2]
    \\
    &= \EE_{\sigma \sim p_\sigma} \EE_{X \sim p_X} \EE_{Y \sim \mathcal{N}(X, \sigma^2\mathbb{I}_{N \times 3})} [\lambda(\sigma) \norm{\cskip Y + \cout \Ft(\cin Y, \cnoise) - x}^2]
    \\
    &= \EE_{\sigma \sim p_\sigma} \EE_{X \sim p_X} \EE_{Y \sim \mathcal{N}(X, \sigma^2\mathbb{I}_{N \times 3})} \left[\lambda(\sigma) \cout^2 \norm{\Ft(\cin Y, \cnoise) - \frac{x - \cskip Y}{\cout}}^2\right]
    \\
    &= \EE_{\sigma \sim p_\sigma} \EE_{X \sim p_X} \EE_{Y \sim \mathcal{N}(X, \sigma^2\mathbb{I}_{N \times 3})} \left[\lambda(\sigma) \cout^2 \norm{\Ft(\cin Y, \cnoise) - F}^2\right]
    \numberthis
    \label{eqn:denoising_loss}
\end{align*}
where:
\begin{align}
    F(y, \sigma) = \frac{x - \cskip y}{\cout}
\end{align}
is the effective training target for the network $\Ft$. We want to normalize $F$ similarly as the network input:
\begin{align}
    \label{eqn:output_norm}
    \EE_{\substack{(i, j) \sim \text{Uniform}(E) \\ \noise \sim \mathcal{N}(0, \mathbb{I}_{N \times 3})}}[\norm{F_i - F_j}^2] = 1
    \ \text{at all noise levels} \ \sigma.
\end{align}
Again, for a fixed $(i, j) \in E$, we have:
\begin{align*}
    \EE_\noise\norm{F_i - F_j}^2 &= \frac{\EE_\noise\norm{(x_i - x_j) - \cskip  (y_i - y_j)}^2}{\cout^2} \\
    &= \frac{\EE_\noise\norm{(1 - \cskip)(x_i - x_j) - \cskip \sigma \cdot (\noise_i - \noise_j)}^2 }{\cout^2}\\
    &= \frac{(1 - \cskip)^2 \norm{x_i - x_j}^2 + \cskip^2 \cdot 6 \sigma^2}{\cout^2}  \numberthis
\end{align*}
and hence:
\begin{align*}
    \EE_{\substack{(i, j) \sim \text{Uniform}(E) \\ \noise \sim \mathcal{N}(0, \mathbb{I}_{N \times 3})}}[\norm{F_i - F_j}^2] &= 1 \\
    \implies \frac{(1 - \cskip)^2 \cdot C + \cskip^2 \cdot 6 \sigma^2}{\cout^2} &= 1
    \\
    \implies \cout^2 = (1 - \cskip)^2 \cdot C + \cskip^2 \cdot 6 \sigma^2
    \numberthis
    \label{eqn:cout_cskip}
\end{align*}
where $C$ was defined above. Now, to minimize $\cout$ to maximize reuse and avoid amplifying network errors, as recommended by \citet{karras2022edm,karras2024edm2}:
\begin{align*}
    \frac{d}{d\cskip} \cout^2 &= 0
    \\
    \implies - 2 (1 - \cskip) \cdot C + 2 \cskip \cdot 6 \sigma^2 &= 0
    \\
    \implies \cskip &=  \frac{C}{C + 6 \sigma^2} \numberthis
\end{align*}
Substituting into \autoref{eqn:cout_cskip}, we get after some routine simplification:
\begin{align}
    \cout &= \sqrt{\frac{C \cdot 6 \sigma^2}{C + 6 \sigma^2}}
\end{align}

From \autoref{eqn:denoising_loss}, we set $\lambda(\sigma) = \frac{1}{\cout^2}$ to normalize the loss at at all noise levels, as in \citet{karras2022edm,karras2024edm2}.
\subsection{Rotational Alignment}
\label{sec:rotational-alignment}

As described in \autoref{alg:alignment}, we use the Kabsch-Umeyama algorithm \citep{kabsch,umeyama} to rotationally align $y$ to $x$ before calling the denoiser.

\begin{algorithm}[h]
\caption{Rotational Alignment with the Kabsch-Umeyama Algorithm}
\label{alg:alignment}
\begin{algorithmic}
\Require Noisy Sample $y \in \R^{N \times 3}$, True Sample $x \in \R^{N \times 3}$.
\Let{$H$}{$x^Ty$} \Comment{$H \in \R^{3 \times 3}$}
\Let{$U, S, V^T$}{$\SVD(H)$} \Comment{ $U, V \in \R^{3 \times 3}$}
\Let{$\mathbf{R}^*$}{$U \text{diag}[1, 1, \det(U)\det(V)]
V^T$}
\State \Return $\mathbf{R}^* \circ y$  
\end{algorithmic}
\end{algorithm}

Note that both $y$ and $x$ are mean-centered to respect translational equivariance:
\begin{align}
    \sum_{i = 1}^N y_i &= \vec{0} \in \R^3 \\
    \sum_{i = 1}^N x_i &= \vec{0} \in \R^3 
\end{align}
so there is no net translation.

\section{Proofs of Theoretical Results}
\label{sec:proofs}

For completeness, we prove the main theoretical results here, as first established by \citet{robbins,miyasawa,saremi2019neural}.

\subsection{The Denoiser Minimizes the Expected Loss}
\label{sec:jump-minimizer-proof}

Here, we prove \autoref{eqn:jump-minimizer}, rewritten here for clarity:
\begin{align}
    \highlight[1]{\denoiser}(\cdot) \equiv 
    \EE[X \ | \ Y = \cdot] = \argmin_{f: \R^{N \times 3} \to \R^{N \times 3}} \EE_{\substack{X \sim p_X, \noise \sim \N(0, \II_{N \times 3}) \\ Y = X + \sigma \noise}}[\norm{f(Y) - X}^2] 
\end{align}

First, we can decompose the loss over the domain $\R^{N \times 3}$ of $Y$:
\begin{align}
    \EE_{\substack{X \sim p_X, \noise \sim \N(0, \II_{N \times 3}) \\ Y = X + \sigma \noise}}[\norm{f(Y) - X}^2] &= \EE_{X \sim p_X, Y \sim p_Y}[\norm{f(Y) - X}^2] \\
    &= \int_{\R^{N \times 3}}\int_{\R^{N \times 3}} \norm{f(y) - x}^2 p_{X, Y}(x, y) dx dy \\
    &= \int_{\R^{N \times 3}}{
        \annotate{
        \int_{\R^{N \times 3}} \norm{f(y) - x}^2 p_{Y | X}(y \ | \ x) p_X(x) dx}{l(f, y)}
    } dy
     \\
    &= \int_{\R^{N \times 3}} l(f, y) dy
\end{align}
where $l(f, y) \geq 0$ for all functions $f$ and inputs $y$.
Hence, any minimizer $f^*$ must minimize the local denoising loss $l(f^*, y)$ at each point $y \in \R^{N \times 3}$.
For a fixed $y \in \R^{N \times 3}$, the loss $l(f, y)$ is convex as a function of $f(y)$. Hence, the global minimizer can be found by finding the critical points of $l(f, y)$ as a function of $f(y)$:
\begin{align}
    \nabla_{f(y)} l(f, y) &= 0
    \\
    \implies  
    \nabla_{f(y)} \int_{\R^{N \times 3}} \norm{f(y) - x}^2 p_{Y | X}(y \ | \ x) p_X(x) dx  &= 0
    \\
    \implies  
    \int_{\R^{N \times 3}} 2 (f^*(y) - x) p_{Y | X}(y \ | \ x) p_X(x) dx  &= 0 
\end{align}
Rearranging:
\begin{align}
    f^*(y) &= \frac{\int_{\R^{N \times 3}} x \ p_{Y | X}(y \ | \ x) p_X(x) dx}{\int_{\R^{N \times 3}} p_{Y | X}(y \ | \ x) p_X(x) dx}
    \\
    &= \frac{\int_{\R^{N \times 3}} x \ p_{Y | X}(y \ | \ x) p_X(x) dx}{p_Y(y)}
    \\
    &= \int_{\R^{N \times 3}} x \ \frac{p_{Y | X}(y \ | \ x) p_X(x)}{p_Y(y)} dx
    \\
    &= \int_{\R^{N \times 3}} x \ p_{X | Y}(x \ | \ y) dx
    \\
    &= \EE[X \ | \ Y = y] 
    \\
    &= \hat{x}(y)
\end{align}
by Bayes' rule. Hence, the denoiser as defined by \autoref{eqn:denoiser-definition} is indeed the minimzer of the denoising loss:
\begin{align}
     \hat{x}(\cdot) \equiv \EE[X \ | \ Y = \cdot] = \argmin_{f: \R^{N \times 3} \to \R^{N \times 3}} \EE_{\substack{X \sim p_X, \noise \sim \N(0, \II_{N \times 3}) \\ Y = X + \sigma \noise}}[\norm{f(Y) - X}^2] 
\end{align}
as claimed.

\subsection{Relating the Score and the Denoiser}
\label{sec:score-denoiser-proof}

Here, we rederive \autoref{eqn:score-denoiser}, relating the score function $\highlight[2]{\nabla \log p_Y}$ and the denoiser $\highlight[1]{\hat{x}}$.

Let $X \sim p_X$ defined over $\R^{N \times 3}$ and $\eta \sim \N(0, \II_{N \times 3})$. Let $Y = X + \sigma \eta$, which means:
\begin{align}
    p_{Y|X}(y \ | \ x) = \N(y; x, \II_{N \times 3}) = \frac{1}{(2\pi\sigma^2)^{\frac{3N}{2}}} \exp\left(-\frac{\norm{y - x}^2}{2\sigma^2} \right)
\end{align}
Then:
\begin{align}
    \EE[X \ | \ Y = y] = y + \sigma^2 \nabla_y \log p_Y(y)
\end{align}
To prove this:
\begin{align}
\nabla_y p_{Y|X}(y \ | \ x) &= -\frac{y - x}{\sigma^2} p_{Y|X}(y \ | \ x) \\
\implies (x - y) p_{Y|X}(y \ | \ x) &= \sigma^2 \nabla_y p_{Y|X}(y \ | \ x) \\
\implies \int_{\R^{N \times 3}} (x - y) p_{Y|X}(y \ | \ x) \ p_X(x) dx &=  \int_{\R^{N \times 3}} \sigma^2 \nabla_y p_{Y|X}(y \ | \ x) \ p_X(x) dx
\end{align}
By Bayes' rule:
\begin{align}
    p_{Y|X}(y \ | \ x) p_X(x) = p_{X,Y}(x, y) = p_{X|Y}(x|y) p_Y(y)
\end{align}
and, by definition of the marginals: \begin{align}
   \int_{\R^{N \times 3}} p_{X,Y}(x, y) dx = p_{Y}(y) 
\end{align}
For the left-hand side, we have:
\begin{align}
\int_{\R^{N \times 3}} (x - y) p_{Y|X}(y \ | \ x) p_X(x) dx &= \int_{\R^{N \times 3}} (x - y) p_{X,Y}(x, y) dx\\
&= \int_{\R^{N \times 3}} x p_{X,Y}(x, y) dx - \int_{\R^{N \times 3}} y p_{X,Y}(x, y) dx \\
&= p_Y(y) \left(\int_{\R^{N \times 3}} x p_{X|Y}(x \ | \ y) dx - y \int_{\R^{N \times 3}} p_{X|Y}(x\ |\ y) dx\right) \\
&= p_Y(y) \left(\mathbb{E}[X \ | \ Y = y] - y \right)
\end{align}
For the right-hand side, we have:
\begin{align}
\sigma^2 \int_{\R^{N \times 3}} \nabla_y p_{Y|X}(y \ | \ x) p_X(x) dx &= \sigma^2 \nabla_y \int_{\R^{N \times 3}} p_{Y|X}(y \ | \ x) p_X(x) dx \\
&= \sigma^2 \nabla_y \int_{\R^{N \times 3}} p_{X,Y}(x, y) dx \\
&= \sigma^2 \nabla_y p_{Y}(y)
\end{align}
Thus,
\begin{align}
p_Y(y) \left(\mathbb{E}[X \ | \ Y = y] - y \right) &= \sigma^2 \nabla_y p_{Y}(y) \\
\implies \mathbb{E}[X \ | \ Y = y] &= y + \sigma^2 \frac{\nabla_y p_{Y}(y)}{p_Y(y)} \\
&= y + \sigma^2 \nabla_y \log p_{Y}(y)
\end{align}
as claimed.
\section{Numerical Solvers for Langevin Dynamics}
\label{sec:baoab}

As mentioned in \autoref{sec:walk-jump}, solving the Stochastic Differential Equation corresponding to Langevin dynamics is often performed numerically. In particular, BAOAB \citep{leimkuhler_rational_2012,leimkuhler_molecular_2015,sachs2017langevin} refers to a `splitting method' that solves the Langevin dynamics SDE by splitting it into three different components labelled by $\mathcal{A}$, $\mathcal{B}$ and $\mathcal{O}$ below:
\begin{align}
    \label{eqn:langevin-baoab}
    dy &= \annotate{v_y dt}{\mathcal{A}}
    \\
    dv_y &= \annotate{M^{-1} \nabla_y \log p_Y(y) dt}{\mathcal{B}} \ -\annotate{\gamma v_y dt + \sqrt{ 2\gamma}M^{-\frac{1}{2}} dB_t}{\mathcal{O}}
\end{align}
where both $y, v_y \in \mathbb{R}^d$. 
This leads to the following update operators:
\begin{align}
    \label{eqn:baoab-operators}
    \mathcal{A}_{\Delta t}
    \begin{bmatrix}
        y \\
        v_y
    \end{bmatrix}
    &=
    \begin{bmatrix}
        y + v_y\Delta t \\
        v_y
    \end{bmatrix}
    \\
    \mathcal{B}_{\Delta t}
    \begin{bmatrix}
        y \\
        v_y
    \end{bmatrix}
    &=
    \begin{bmatrix}
        y \\
        v_y + M^{-1}\nabla_y \log p_Y(y) \Delta t
    \end{bmatrix}
    \\
    \mathcal{O}_{\Delta t}
    \begin{bmatrix}
        y \\
        v_y
    \end{bmatrix}
    &=
    \begin{bmatrix}
        y \\
        e^{-\gamma \Delta t} v_y + M^{-\frac{1}{2}}\sqrt{1 - e^{-2\gamma \Delta t}} B
    \end{bmatrix}
\end{align}
where $B \sim \N(0, \mathbb{I}_d)$ is resampled every iteration. As highlighted by \citet{gromacs-baoab}, the $\mathcal{A}$ and $\mathcal{B}$ updates are obtained by simply discretizing the updates highlighted in \autoref{eqn:langevin-baoab} by the Euler method. The $\mathcal{O}$ update refers to a explicit solution of the Ornstein-Uhlenbeck process, which we rederive for completeness in \autoref{sec:ou-solve}.

Finally, the iterates of the BAOAB algorithm are given by a composition of these update steps, matching the name of the method:
\begin{align}
    \label{eqn:baoab-update}
    \begin{bmatrix}
        y^{(t + 1)} \\
        v_y^{(t  + 1)}
    \end{bmatrix}
    = 
    \mathcal{B}_{\frac{\Delta t}{2}}
    \mathcal{A}_{\frac{\Delta t}{2}}
    \mathcal{O}_{\Delta t}
    \mathcal{A}_{\frac{\Delta t}{2}}
    \mathcal{B}_{\frac{\Delta t}{2}}
    \begin{bmatrix}
        y^{(t)} \\
        v_y^{(t)}
    \end{bmatrix}
\end{align}

\section{The Ornstein-Uhlenbeck Process}
\label{sec:ou-solve}

For completeness, we discuss the distributional solution of the Ornstein-Uhlenbeck process, taken directly from the excellent \citet{leimkuhler_molecular_2015}.
In one dimension, the Ornstein-Uhlenbeck Process corresponds to the following Stochastic Differential Equation (SDE): 
\begin{align}
    dv_y = -\gamma v_y dt + \sqrt{2\gamma}M^{-\frac{1}{2}} dB_t
\end{align}
Multiplying both sides by the integrating factor $e^{\gamma t}$:
\begin{align}
    e^{\gamma t} dv_y &= -\gamma e^{\gamma t} (v_y dt + e^{\gamma t} \sqrt{2\gamma}M^{-\frac{1}{2}} dB_t
    \\
    \implies
    e^{\gamma t} (dv_y + \gamma v_y dt) &= e^{\gamma t} \sqrt{2\gamma}M^{-\frac{1}{2}} dB_t
\end{align}
and identifying:
\begin{align}
    e^{\gamma t}(dv_y + \gamma v_ydt) 
    = d(e^{\gamma t}v_y)
\end{align}
We get after integrating from $t_1$ to $t_2$, which represent two adjacent time steps in our discretized grid:
\begin{align}
    d(e^{\gamma t}v_y) &= e^{\gamma t} \sqrt{2\gamma}M^{-\frac{1}{2}} dB_t
    \\
    \implies 
    \int_{t_1}^{t_2} d(e^{\gamma t}v_y) &= \int_{t_1}^{t_2} e^{\gamma t} \sqrt{2\gamma}M^{-\frac{1}{2}} dB_t
    \\
    \implies 
    e^{\gamma t_2}v_y(t_2) - e^{\gamma t_1}v_y(t_1) &= \sqrt{2\gamma}M^{-\frac{1}{2}} \int_{t_1}^{t_2} e^{\gamma t}  dB_t
\end{align}
Now, for a Wiener process $B_t$, if $g(t)$ is a deterministic function, $\int_{t_1}^{t_2} g(t) dB_t$ is distributed as 
$\N\left(0, \int_{t_1}^{t_2} g(t)^2 dt \right)$ by It\^{o}'s integral. Thus, applying this result to $g(t) = e^{\gamma t}$, we get:
\begin{align}
    e^{\gamma t_2}v_y(t_2) - e^{\gamma t_1}v_y(t_1) &= \sqrt{2\gamma}M^{-\frac{1}{2}} \N\left(0, \frac{e^{2 \gamma t_2} - e^{2 \gamma t_1}}{2\gamma}\right)
    \\
    \implies
    v_y(t_2) &= e^{-\gamma(t_2 - t_1)}v_y(t_1) + \sqrt{2\gamma}M^{-\frac{1}{2}}e^{-\gamma t_2} \N\left(0, \frac{e^{2 \gamma t_2} - e^{2 \gamma t_1}}{2\gamma}\right)    \\
    &= e^{-\gamma(t_2 - t_1)}v_y(t_1) + \sqrt{2\gamma}M^{-\frac{1}{2}} 
    \sqrt{\frac{1 - e^{2 \gamma (t_1 - t_2)}}{2\gamma}}
    \N\left(0, 1\right)
    \\
    &= e^{-\gamma(t_2 - t_1)}v_y(t_1) + M^{-\frac{1}{2}} 
    \sqrt{1 - e^{2 \gamma (t_1 - t_2)}}
    \N\left(0, 1\right)
\end{align}
In the ${N \times 3}$ dimensional case, as the Wiener processes are all independent of each other, we directly get:
\begin{align}
    v_y(t_2) &= e^{-\gamma(t_2 - t_1)}v_y(t_1) + M^{-\frac{1}{2}} 
    \sqrt{1 - e^{2 \gamma (t_1 - t_2)}}
    \N\left(0, \II_{N \times 3} \right)
\end{align}

Setting $\Delta t = t_2 - t_1$, we get the form of the $\mathcal{O}$ operator (\autoref{eqn:baoab-operators}) of the BAOAB integrator in \autoref{sec:baoab}.
\section{Parallelizing Sampling with Multiple Independent Chains}
\label{sec:parallel-sampling}

Our sampling strategy batches peptides in order to increase throughput. Another potential method to increase throughput (but which we did not employ for the results in this paper) is to sample multiple chains in parallel.

This can be done by initializing multiple chains: $y^{(0)}_1, \ldots, y^{(0)}_{N_{\text{ch}}}$, where:
\begin{align}
    y^{(0)}_{\text{ch}} = x^{(0)} + \sigma \noise^{(0)}_{\text{ch}}
\end{align}
where $\noise^{(0)}_{\text{ch}} \overset{\mathrm{iid}}{\sim}  \mathcal{N}(0, \mathbb{I}_{N \times 3})$ for $\text{ch} = 1, \ldots, N_\text{ch}$ are all independent of each other. Then, the chains can be evolved independently withindependent walk steps (\autoref{eqn:langevin}) and denoised with independent jump steps (\autoref{eqn:score-denoiser}). This independence allows batching over the $y^{(t)}_{\text{ch}}$ over all chains $\text{ch}$ at each iteration $t$. 

Note that at $t = 0$, the chains are correlated as they are all initialized from the same $x^{(0)}$. However, if the number of samples per chain is large enough, the chains are no longer correlated, as they have now mixed into the stationary distribution.
\end{document}